\documentclass[12pt,preprint]{aastex}
\usepackage{rotating,amsfonts,epsfig, enumerate}
\begin{document}

\bibliographystyle{apj}
                                                              
\title{A New Determination of the High Redshift Type Ia Supernova Rates with the Hubble Space Telescope Advanced Camera for Surveys.\altaffilmark{1}}

\author{N.~Kuznetsova\altaffilmark{2,3},
K.~Barbary\altaffilmark{2,4},
B.~Connolly\altaffilmark{5},
A.~G.~Kim\altaffilmark{2},
R.~Pain\altaffilmark{6},
N.~A.~Roe\altaffilmark{2},
G.~Aldering\altaffilmark{2},
R.~Amanullah\altaffilmark{7},
K.~Dawson\altaffilmark{2},
M.~Doi\altaffilmark{8}
V.~Fadeyev\altaffilmark{9},
A.~S.~Fruchter\altaffilmark{10},
R.~Gibbons\altaffilmark{11},
G.~Goldhaber\altaffilmark{2,4},
A.~Goobar\altaffilmark{12},
A.~Gude\altaffilmark{4},
R.~A.~Knop\altaffilmark{11},
M.~Kowalski\altaffilmark{13},
C.~Lidman\altaffilmark{14},
T.~Morokuma\altaffilmark{15}
J.~Meyers\altaffilmark{2,4},
S.~Perlmutter\altaffilmark{2,4},
D.~Rubin\altaffilmark{2,4},
D.~J.~Schlegel\altaffilmark{2},
A.~L.~Spadafora\altaffilmark{2},
V.~Stanishev\altaffilmark{12},
M.~Strovink\altaffilmark{2,4},
N.~Suzuki\altaffilmark{2},
L.~Wang\altaffilmark{16},
N.~Yasuda\altaffilmark{17} (Supernova Cosmology Project)
}

\altaffiltext{1}{Based on observations with the NASA/ESA \emph{Hubble Space Telescope}, obtained at the Space Telescope Science Institute, which is operated by AURA, Inc., under NASA contract NAS 5-26555, under programs GO-9583, GO-9425, GO-9727, and GO-9728.}
\altaffiltext{2}{E. O. Lawrence Berkeley National Laboratory, 1 Cyclotron Rd., Berkeley, CA 94720, USA }
\altaffiltext{3}{Current address: Physics Department, Hamilton College, Clinton, NY 13323, USA}
\altaffiltext{4}{Department of Physics, University of California Berkeley, Berkeley, CA 94720, USA}
\altaffiltext{5}{Department of Physics, Columbia University, New York, NY 10027, USA}
\altaffiltext{6}{LPNHE, CNRS-IN2P3, University of Paris VI \& VII, Paris, France }
\altaffiltext{7}{The Space Sciences Laboratory, University of California Berkeley, Berkeley, CA 94720, USA}
\altaffiltext{8}{Institute of Astronomy, School of Science, University of Tokyo, Mitaka, Tokyo, 181-0015, Japan}
\altaffiltext{9}{Department of Physics, University of California Santa Cruz, Santa Cruz, CA 95064, USA}
\altaffiltext{10}{Space Telescope Science Institute, 3700 San Martin Drive, Baltimore, MD 21218, USA}
\altaffiltext{11}{Department of Physics and Astronomy, Vanderbilt University, Nashville, TN 37240, USA}
\altaffiltext{12}{Department of Physics, Stockholm University, Albanova University Center, S-106 91 Stockholm, Sweden}
\altaffiltext{13}{Humboldt Universit\"{a}t Institut f\"{u}r Physik, Newtonstrasse 15, Berlin 12489, Germany}
\altaffiltext{14}{European Southern Observatory, Alonso de Cordova 3107, Vitacura, Casilla 19001, Santiago 19, Chile}
\altaffiltext{15}{Optical and Infrared Astronomy Division, National Astronomical Observatory of Japan, Mitaka, Tokyo, 181-8588, Japan}
\altaffiltext{16}{Department of Physics, Texas A\&M University, College Station, TX 77843, USA}
\altaffiltext{17}{Institute for Cosmic Ray Research, University of Tokyo, Kashiwa, 277 8582, Japan}

\email{nvkuznetsova@lbl.gov}



\begin{abstract}
We present a new measurement of the volumetric rate of Type Ia supernova  up 
to a redshift of 1.7, using the Hubble Space Telescope (HST) GOODS data combined 
with an additional HST dataset covering the North GOODS field collected in 2004. 
We employ a novel technique that does not require spectroscopic data  
for identifying Type Ia supernovae
(although spectroscopic measurements of redshifts are used for over  
half the sample); instead we employ a Bayesian approach using only  
photometric
data to calculate the probability that an object is a Type Ia supernova.
This Bayesian technique can easily be modified 
to incorporate improved priors on supernova properties, and it is well-suited for future 
high-statistics supernovae searches in which spectroscopic follow up of all candidates 
will be impractical.  Here, the method is validated on both ground- and space-based supernova data having some spectroscopic follow 
up. We combine our volumetric rate measurements with low redshift supernova data, 
and fit to a number of possible models for the evolution of the Type Ia supernova rate as 
a function of redshift. The data do not distinguish between a flat rate at 
redshift $>$ 0.5 and a previously proposed model, in which the Type Ia rate peaks at redshift 
$\sim$ 1 due to a significant delay from star-formation to the supernova explosion.  Except for the highest redshifts, where the signal to noise ratio is  
generally too low to apply this technique,
this approach yields smaller or comparable uncertainties than previous work.
\end{abstract}

\section{Introduction}
The empirical evidence for the existence of dark energy came from observations of
Type Ia supernovae~\citep{bib:riess, bib:P99, bib:P03}, which 
are believed to arise from the thermonuclear explosion of a progenitor white dwarf 
after it approaches the Chandrasekhar mass limit~\citep{bib:chan}.  
However, the physics
of Type Ia supernova production is not well understood.  
The two most plausible scenarios for the white dwarf to accrete the
necessary mass are the single degenerate case, where the white dwarf is 
located in a binary system; and the double
degenerate case, where two white dwarfs merge.  The 
Type Ia supernova rate is correlated with the star formation history (SFH), and
thus a measurement of the rate as a function of redshift helps constrain the possible type 
Ia progenitor models.

In addition to its importance for understanding Type Ia supernovae as astronomical objects, 
a good grasp of the Type Ia supernova rate to high redshifts is important
for the next generation of proposed space-based supernova cosmology experiments,
such as SNAP~\citep{bib:snap}.
It is therefore of great practical interest to determine the rate of Type Ia supernovae
at redshifts $>$ 1.

The subject of Type Ia supernova rates has been addressed by many authors
in the past. 
Existing rate measurements have been mostly limited to redshift ranges $<$ 1:
the results of~\cite{bib:cappellaro},~\cite{bib:hardin},~\cite{bib:madgwick}, and~\cite{bib:blanc}
measure the rates at redshifts $\leq$ $\sim$0.1;~\cite{bib:neill},~\cite{bib:tonry}, and~\cite{bib:pain},
at intermediate redshifts of 0.47, 0.50, and 0.55, respectively; and~\cite{bib:barris},
up to a redshift of 0.75.  The only published measurement of the rates
at redshifts $>$ 1 is that of~\cite{bib:dahlen}, who analyzed the GOODS dataset.  

There are several important 
differences that distinguish our work from that of~\cite{bib:dahlen}.
First, we augment the GOODS sample with the HST data collected during 
the Spring-Summer 2004 high redshift supernova searches.
Second, our methods of calculating the control time (the time during which
a supernova search is potentially capable of finding supernova candidates)
and the efficiency to identify a supernova are based on a detailed Monte Carlo
simulation technique using a library of supernova templates.
Third, we adopt a novel approach
to typing supernovae, using photometric data and a 
Bayesian probability method described in~\cite{bib:ourpaper}. 
The Bayesian technique is able to perform classification
using only photometric data, and therefore does not require
spectroscopic follow up.  Optionally, photometric or 
spectroscopic redshifts can be used to improve the classification accuracy.
Our initial requirements on potential supernova candidates 
are more stringent in terms of the number of points on the light curve
and the signal to noise of those points
than those of~\cite{bib:dahlen}; thus some of the candidates they
identified will fail our cuts.  
However, we are able to reliably separate Type Ia supernova 
from other supernovae types based on their Bayesian probability, with
an efficiency that is readily quantifiable, thus allowing 
us to use larger data samples.  
Our approach therefore avoids the problems
that arise in estimating the efficiency for the decision to schedule 
spectroscopic follow up based on a potentially low signal-to-noise
initial detection.

The Bayesian classification technique uses photometric data, and
does not require any spectroscopic followup.  This is an advantage for 
future large-area surveys  (such as the Dark Energy Survey,
Pan-STARRS, and LSST) that will discover thousands of supernova candidates,
but are unlikely  to be able to obtain spectroscopic data for all of them, to distinguish
Type Ia supernovae from core collapse supernovae and other variable objects.
The technique described here can be
considered a prototype of the kind of analysis that could be performed on
these future large data sets to identify Type Ia supernovae for cosmological studies.
There is a clear trade-off involved in using photometric measurements alone:
if the quality of the photometric data is poor, then
the efficiency of this technique to identify Type Ia supernovae
is reduced; on the other hand, this technique enables larger samples
of Type Ia from imaging surveys to be identified for cosmological studies,
without the need for time-consuming spectroscopic follow up.

Note that although the method is able to perform the supernova typing with photometric data alone (\emph{i.e.}, it does not require spectroscopic data, either 
redshifts or types), it is certainly able to use the extra information that is
available, and in fact 70\% of the supernova
candidates discussed in the present work have redshifts which were  
obtained spectroscopically.
It is also worth noting that while in this paper we only analyze the Type Ia supernova rates,
the  Bayesian classification technique can be used to classify other types as well, 
making it possible to measure the rates of non-Type Ia supernovae in a similar fashion.  
These analyses will be presented in future publications.

The paper is organized as follows.  In section~\ref{sec:data} we describe
the data samples used in the analysis.  In section~\ref{sec:candselection}
we describe the supernova candidate selection and typing process.
In section 4 we calculate the control time, survey area, 
and search efficiency, and determine the volumetric Type Ia
supernova rate from our data sample.
A comparison of the rates with those reported in the literature is given 
in section~\ref{sec:comp}, and fits of the rates to different models
relating the Type Ia supernova rates to the SFH
are given in section~\ref{sec:sf}.
A summary is given in section~\ref{sec:concl}.

\section{Data Sample}
\label{sec:data}
\begin{figure}[!htb]
 \begin{center}
$\begin{array}{c@{\hspace{0.0in}}c}
\epsfxsize=3.in\epsfysize=3.2in\epsffile{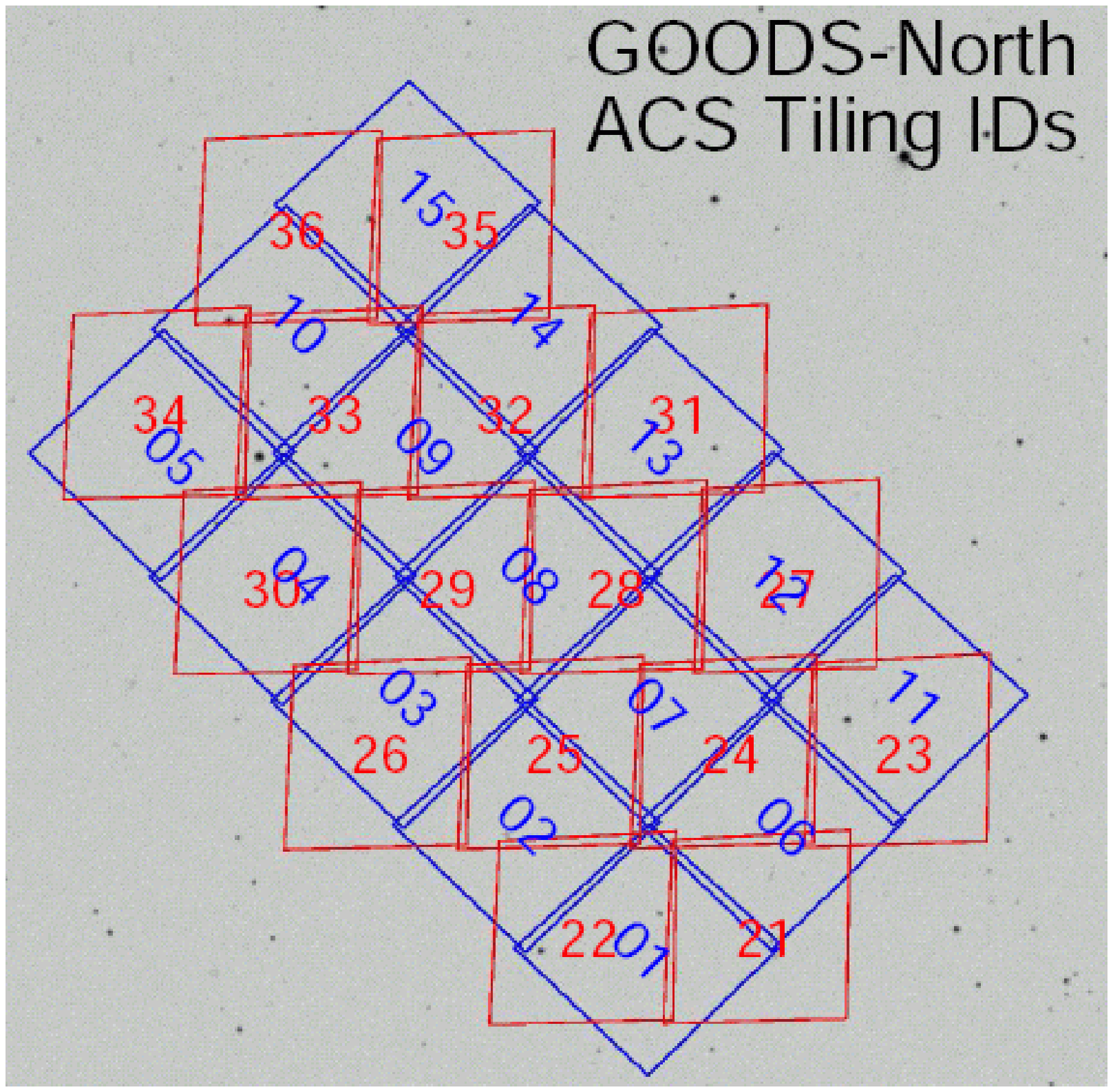} & 
\epsfxsize=3.in\epsfysize=3.2in\epsffile{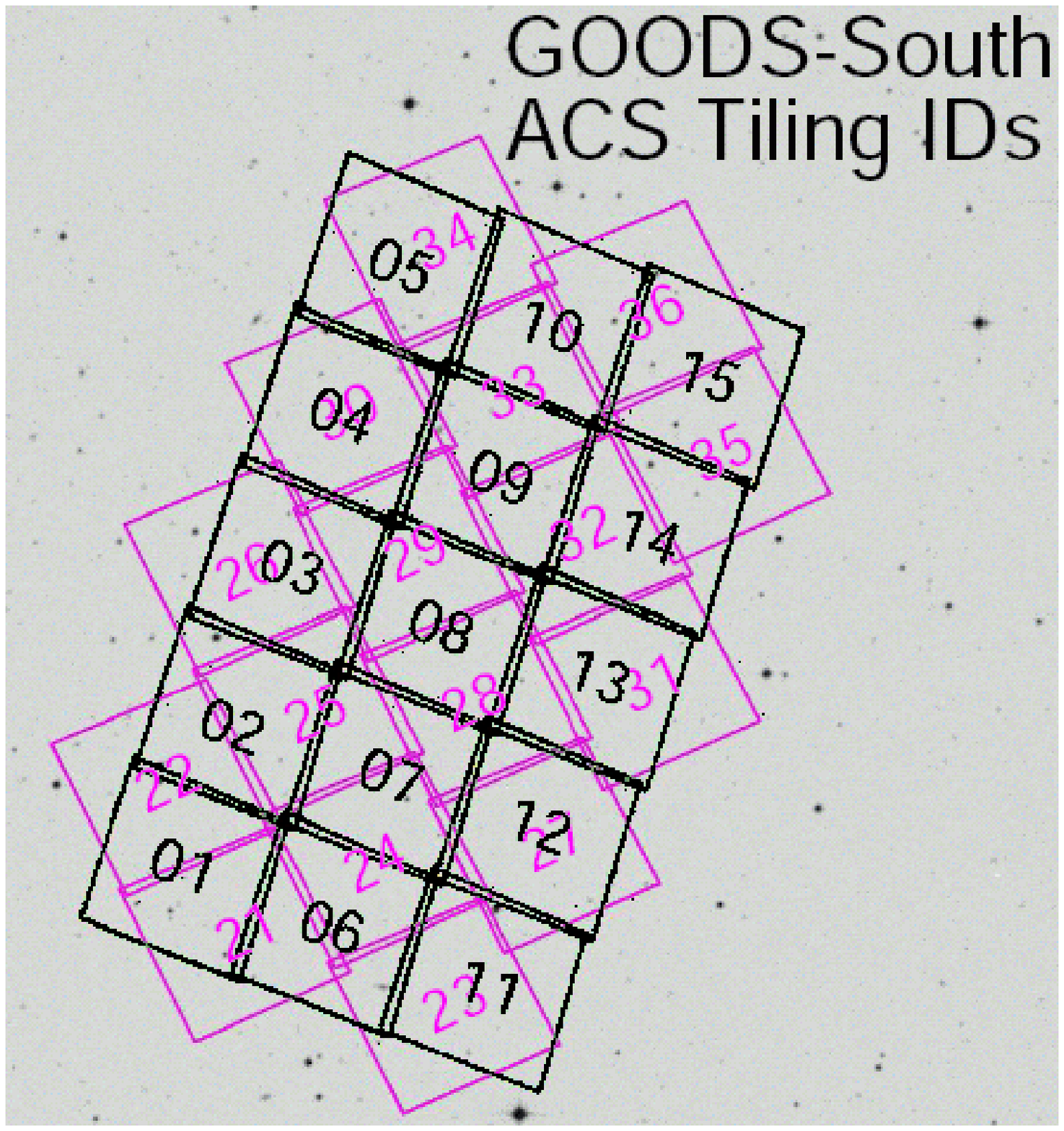}
\end{array}$
  \end{center}
\caption[]
{\label{fig:goods}
The North (left) and South (right) GOODS fields.  
The fields are subdivided into tiles, which are shown (along with their ID numbers) 
on the figures.  The size of a single tile is $\sim$11.5 sq. arcmin.
}
\end{figure}

For this analysis, we use the Hubble Space Telescope (HST) 
GOODS dataset collected in 2002-2003~\citep{bib:goods2, bib:goods1, bib:goods0}.  
In addition to the GOODS data, we use an HST sample collected
in the Spring-Summer of 2004, which hereafter we will call the 2004 ACS sample.
The GOODS dataset consists of five epochs (data taking periods), separated by approximately 
45 observer-frame days.  The GOODS data used for this analysis  were taken in
two HST ACS filter bands: F775W (centered at 775 nm) and F850LP (centered at 850 nm)\footnote[1]{The ACS filter transmission curves
are available at http://acs.pha.jhu.edu/instrument/filters/.}.
Each F850LP image consists of four exposures; and each F775W image, of two.
The GOODS survey includes two fields, GOODS North and GOODS South,
and covers approximately 320 square arcminutes.  The fields are sub-divided into
smaller ``tiles'' that correspond to single ACS pointings 
(typically 15 or 16), as shown in Fig.~\ref{fig:goods}.

The 2004 ACS supernova dataset covers only the GOODS North field, with 
the same tiling as that of the GOODS North dataset.  It consists of 4 epochs
separated by approximately 45 observer-frame days.
The data in this sample were taken in two HST ACS passbands:
F775W and F850LP, with one exposure for every F775W image and four
for every F850LP image.  Two teams (PI Perlmutter and PI Riess) shared this data searching for 
supernovae in alternate visits;~\cite{bib:riess06} have published the 
results for the supernovae that were discovered in their team's visits.

For convenience, a summary of the datasets used
is given in Table~\ref{tab:data}.
\begin{table}[htbp]
  \begin{center}
  \begin{minipage}{\textwidth}
    \begin{tabular}{|c|c|c|c|c|c|c|}
      \hline\hline
       Epoch & Filter & Exp.time (s) &  Filter & Exp. time (s) &  \# Tiles & Taken On \\
     \hline\hline
      \multicolumn{7}{|c|}{GOODS South}  \\
\hline
 	1 &	F775W&	1040 &	F850LP  & 2120 &  15 & 7/31 - 8/4 (2002)\\
 	2 &	F775W &	1040 &	F850LP  & 2120 &  16 & 9/19 - 9/22 (2002)\\
 	3 &	F775W &	1040 &	F850LP  & 2120 &  15 & 10/31 - 11/3 (2002)\\
 	4 &	F775W &	1040 &	F850LP  & 2120 &  16 & 12/19 - 22 (2002)\\
	5 &	F775W &	1040 &	F850LP  & 2120 &  15 & 2/1 - 2/5 (2003)\\
     \hline\hline
      \multicolumn{7}{|c|}{GOODS North}  \\
      \hline
        1  &	F775W &	1120 & F850LP   & 2400 & 14  &  11/21 - 11/22 (2002)\\
 	2 &	F775W &	1000\footnote{except for tile 30 (1060)} & F850LP & 2120 & 17 & 1/2 - 1/4 (2003)\\
 	3 &	F775W &	960   & F850LP  & 2060 & 16 & 2/20 - 2/23 (2003) \\
 	4 &	F775W &	960   & F850LP  & 2000 & 14 &  4/3 - 4/6 (2003)\\
 	5 &	F775W &	960   & F850LP  & 2080 & 15 & 5/21 - 5/25 (2003)\\
     \hline\hline
      \multicolumn{7}{|c|}{2004 ACS Sample (GOODS North tiles)}  \\
      \hline 
       1 &  F775W &  400  &  F850LP & 1600 & 15  &     4/2 - 4/4 (2004) \\
       2 &  F775W &  400  &  F850LP & 1600 & 15  &     5/20 - 5/23 (2004) \\
       3 &  F775W &  400  &  F850LP & 1600 & 15  &     7/9 - 7/10  (2004) \\
       4 &  F775W &  400  &  F850LP & 1600 & 15  &    8/26 - 8/28 (2004)\\
     \hline\hline
   \end{tabular}
   \end{minipage}
 \end{center}
\caption[]
{\label{tab:data}
A summary of the datasets used in this analysis, listing the data taking epochs,
the filters, the exposure times of the combined exposures (in seconds), 
the number of GOODS field tiles, and the dates when the data were taken.
}
\end{table}

It is worth emphasizing that we are using photometric information from only 
two filter bands, providing one color measurement.  
The GOODS dataset has been analyzed before, and 
13 out of 42
supernovae found were spectroscopically typed~\citep{bib:riess2, bib:strolger}.
For the 2004 ACS sample, however, the spectroscopic information is
available only for a small fraction of the candidates.
We treat both GOODS and 2004 ACS datasets in a consistent fashion, using
photometric information only for typing supernovae (note that we still
use spectroscopically determined redshifts where available).  
This allows more data to be searched
and more supernovae to be found, but at the expense of neglecting 
spectroscopic information for the candidates where it is available.
In section~\ref{sec:typeia} we discuss
in detail the resulting supernova candidate count.

We start with the data that have been flat-fielded and gain-corrected
by the HST pipeline, and use MultiDrizzle~\citep{bib:drizzle} to 
perform cosmic ray rejection and to combine dithered observations.
The parameters of the drizzling process include a  ``square'' kernel, with a pixel fraction of 0.66
and a pixel scale of 1.0.  The drizzling 
combines the multiple individual pointings.
Drizzling is ineffective for the cosmic ray rejection for 
the F775W data from the 2004 ACS sample since they contain
only a single exposure for each GOODS North tile.  We therefore
use a morphological cosmic ray rejection package~\citep{bib:lacosmics} to 
create images with identifiable objects, thus allowing us to generate 
the geometrical transformations between images; however, the original
images are used for extracting photometric information (after verifying 
that no cosmic rays landed directly at the location of the supernova candidates).

Supernovae are identified by subtracting a reference image from each 
of the HST search epochs.
We create four distinct samples summarized in Table~\ref{tab:samples}, which 
we use for identifying and performing simple aperture photometry on 
the supernova candidates in each of the five epochs in the GOODS dataset
and each of the four epochs in the 2004 ACS dataset.  
To obtain the multi-epoch photometry
for the GOODS North data (sample \#1), we combine all four epochs of the 2004 ACS sample
and then subtract these data from each of the  five North GOODS epochs in turn.
Combining multiple epochs for the reference image allows us to create deeper
resulting data, which is important for extracting supernovae with
the best possible signal-to-noise ratio (SNR).
For sample \#2, we combine the entire North GOODS sample and subtract these data
from each of the four 2004 ACS epochs in turn.  
Because the GOODS and 2004 ACS data were taken with a time separation of approximately 
a year, these samples should be sensitive to the supernovae that were both on the rise and
on the decline during the GOODS and 2004 ACS data taking period for samples \#1 and \#2,
respectively.
For the GOODS South sample, however, we do not have any additional datasets, and
are thus forced to separate the sample into two.  This is the reason 
the three initial data samples (GOODS North and South and the 2004 ACS data set)
result in four search samples.
We combine South epochs 4 and 5 for sample \#3, and epoch 1 and 2 for sample \#4;
we then subtract the two combined samples separately from each of the five 
South GOODS epochs.
\begin{table}[htbp]
  \begin{center}
    \begin{tabular}{|c|c|c|}
    \hline\hline
      Sample \#           & Reference Dataset                    &  Supernova Search        \\
     \hline\hline
          1               & Combined 2004 ACS data (4 epochs)  & Individual North GOODS epochs\\
          2               & Combined North GOODS data (5 epochs) & Individual 2004 ACS dataset epochs \\
          3               & Epochs 4+5 of the South GOODS data   & Individual South GOODS epochs\\
          4               & Epochs 1+2 of the South GOODS data   & Individual South GOODS epochs\\
     \hline\hline
   \end{tabular}
 \end{center}
\caption[]
{\label{tab:samples}
The samples used in our supernova search.  To identify and extract photometry for supernova candidates,
we subtract the data listed in column 2 from the data listed in column 3.  Note 
that sample \#2 has the deepest references.}
\end{table}
If a supernova candidate has been found in both samples \#3 and \#4, we consider it
belonging to the sample in which it had an epoch with the largest 
SNR.  This avoids any possible double-counting of 
the candidates for the GOODS South data.

\section{The Supernova Candidate Selection and Typing}
\label{sec:candselection}
The search for supernova candidates and their subsequent
typing as Ia's is a 3-stage process.  We will briefly describe them
below, and then in detail in sections~\ref{sec:initialsearch}-\ref{sec:typeia}.
\begin{enumerate}
\item
First, potential supernova candidates in individual epochs are identified
by the software that is used to subtract the supernova
search data from the reference data.  
The initial candidate selection is done using the F850LP data only
because it suffers less from cosmic ray contamination,
and because F850LP covers supernovae at redshifts up to $\sim$1.5.
The initial supernova selection  is primarily directed toward
reducing the number of false positives resulting from various image processing artifacts
and residual cosmic ray contamination.
It is followed by a manual scan to reject any obvious remaining cosmic rays and
image processing artifacts.  
Note that both sources of false detections have specific signatures that 
real supernovae do not have; this selection therefore is not expected
to reduce the number of real supernovae in the sample.
This stage is described in detail in section~\ref{sec:initialsearch}.
\item
For the candidates on individual epochs that pass the first stage of the selection process,
we extract the photometric information at the candidate locations
in the multi-epoch F850LP and F775W data.
We then select candidates with reasonably well-measured light curves 
by requiring that the candidate's SNR in the 
subtracted data (in both filters) be greater than 2 for at least 3 search epochs,
including at least two with a SNR greater than 3.
At the end of this stage, we are left with the majority of candidates that
are presumed to be supernovae of some type, as well as some candidates
that cannot be modeled as any known supernova type.  
This stage is described in detail in section~\ref{sec:multibandsearch}.
\item
The final step applies a Bayesian likelihood technique that assigns each candidate that passed steps 1 and 2 
a probability to be a Type Ia supernova based on the multi-epoch data in both filters.
This stage is described in detail in section~\ref{sec:typeia}.
\end{enumerate}  
For convenience, we summarize the selection process in Table~\ref{tab:selection}.
We now describe each of the selection stages in detail.
\begin{table}[htbp]
  \begin{center}
    \begin{tabular}{|c|c|c|}
    \hline\hline
         Selection Stage           & Data Used              & Cuts Applied\\
     \hline\hline
                          &                                            &  SNR$_{\rm exposure}$ $>$ 3 in 4 exposures\\
          1               & F850LP                                     &  SNR consistency in 3 out of 4 exposures\\
                          & single (discovery) epoch                   &  Percent increase $\geq$ 15\% in combined exposures \\
                          &                                            &  shape cuts in combined exposures\\
	  \hline
          2               & F850LP, F775W, all epochs  & $\geq$ 3 epochs with S/N $>$ 2 \\
                          &                            & (including $\geq$ 2 epochs with S/N $>$ 3) \\
	  \hline
          3               & F850LP, F775W, all epochs  &  Bayesian Type Ia classification \\
     \hline\hline
   \end{tabular}
 \end{center}
\caption[]
{\label{tab:selection}
A summary of the Type Ia supernova selection and typing process.  The meaning of the cuts
is explained in the text describing the corresponding stages.}
\end{table}

\subsection{Stage 1: Single Epoch Supernova Candidate Selection}
\label{sec:initialsearch}
In the first step of the supernova search, we search for supernova candidates
in the individual epochs of the F850LP data by looking for signals in the reference-subtracted search 
images.  The reference image is the same for each exposure
(recall that each F850LP image consists of four exposures, each
with the same exposure time).  We use aperture photometry 
with a radius of 3 pixels, where the pixel scale is 0.03'' (after drizzling).
This choice of the aperture optimizes the SNR of supernova candidates.
We verified that the photometric extraction procedure is working 
well by creating ``fake'' supernovae, as described later in this section,
 and comparing their input and output magnitudes;
they agree at the sub-percent level.  The procedure for identifying supernovae is as follows.
\renewcommand{\labelenumi}{\roman{enumi}.}
\begin{enumerate}
\item
Subtracting the combined (drizzled) exposures of the search data from the (drizzled) reference data,
we require that:
 \begin{itemize}
  \item
   The absolute value of the flux within the supernova candidate's aperture 
in the subtracted data divided by the flux in the reference data (the ``percent
increase'' variable) be $\geq$ 15\%.
   \item
   The candidate's shape in the subtracted data must be consistent with a point source:
we require that the candidate's FWHM in both $x$- and $y$- directions 
be $<$ 4 pixels, and that the absolute value of its normalized $xy$ moment be $<$ 0.5 pixels.
  \end{itemize}
\item
Next, to eliminate false detections resulting from cosmic rays, we do
the following:
 \begin{itemize}
 \item
We consider the four individual exposures of the search 
images.  The SNR measured 
for a supernova candidate in each of these exposures (SNR$_{\rm exposure}$)
should be at least 3.  A false positive resulting from cosmic rays will likely
not be present in every individual exposure.
 \item
We then subtract each of the individual exposures from the reference
image at the location of the supernova candidate 
and compare the signal to the quadratic sum of the noise.  The difference
in these SNRs between the exposures must be $<$ 3 for at least 3 out of 4 exposures.  We are thus allowing one
(and only one) of the four exposures of the search image
to be contaminated by a cosmic ray.
 \end{itemize}
\end{enumerate}

These cuts eliminate close to 90\% of false detections
(\emph{i.e.}, the number of detections decreases from $\sim$100
per single tile (see Fig.~\ref{fig:goods}) to $\sim$10).
Obvious image processing artifacts or cosmic rays that manage to 
pass these cuts are rejected by manual screening
(typically, there would be a few such candidates per tile, 
mostly image processing artifacts), with any
questionable candidates left in the sample.  
The efficiency of the manual
scan has been checked using a sample of $\sim$100 fake supernovae,
generated as described below, and 100\% were correctly identified.
The preliminary selection flags any variable objects -- supernovae of various types, as
well as active galactic nuclei (AGNs), \emph{etc}.  In section~\ref{sec:typeia},
we describe our approach to selecting Type Ia supernovae from the sample.

In order to measure the efficiency of the selection, we used a Monte Carlo simulation
that puts fake supernovae on real F850LP images.  
Fake supernovae were also used to develop the selection cuts listed above
in an unbiased way.  The technique follows the
approach outlined in~\cite{bib:pain1} and works as follows.

First, we run SExtractor~\citep{bib:sex} v2.3 on the search images
that have been combined, or drizzled, together from the individual exposures.
We do this for a number of both North and South GOODS tiles.
Using SExtractor's classification
of objects as galaxies and stars, we create a list of the galaxy positions 
on the image.  Because in our analysis we are ignoring candidates near image
edges, the galaxies located within 2 galaxy full widths at half maximum (also 
determined by SExtractor) from the image boundaries are discarded.  
The fake supernova that is to be put on the image is randomly 
assigned a magnitude that is drawn from a flat distribution between 23 to 30.
The supernova's position is drawn from a Gaussian distribution
with half the galaxy's full width at half maximum as the 
standard deviation and centered on the galaxy's nominal center.
We then use STSDAS\footnote[2]{STSDAS and PyRAF are products of the Space 
Telescope Science Institute, which is operated by AURA for NASA}
\emph{tranback} function to convert the fake supernova positions on the
drizzled images into coordinates on the raw individual exposures.
Fake supernovae themselves are created using the TinyTim software~\citep{bib:tt}, for 
the ACS WFC1 camera, in filter F850LP.  The fake supernova signal,
combined with a noise generated using a Poisson distribution with the signal's mean
for each pixel, 
is added onto the input exposures, which are subsequently processed in exactly the same way as real data 
are.

\begin{figure}[!htb]
\begin{center}
$\begin{array}{c@{\hspace{0.0in}}c}
\epsfxsize=3.0in\epsffile{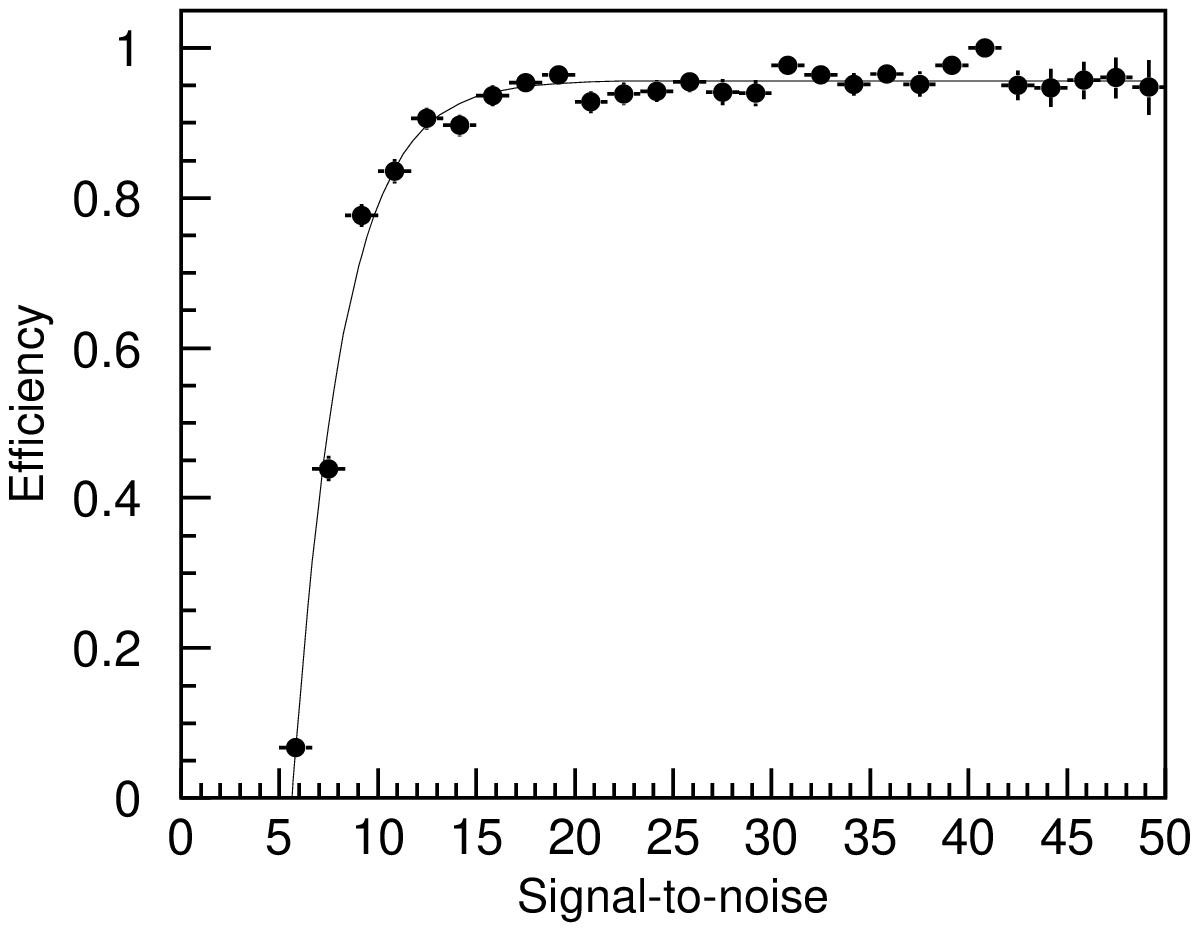} & 
\epsfxsize=3.0in\epsffile{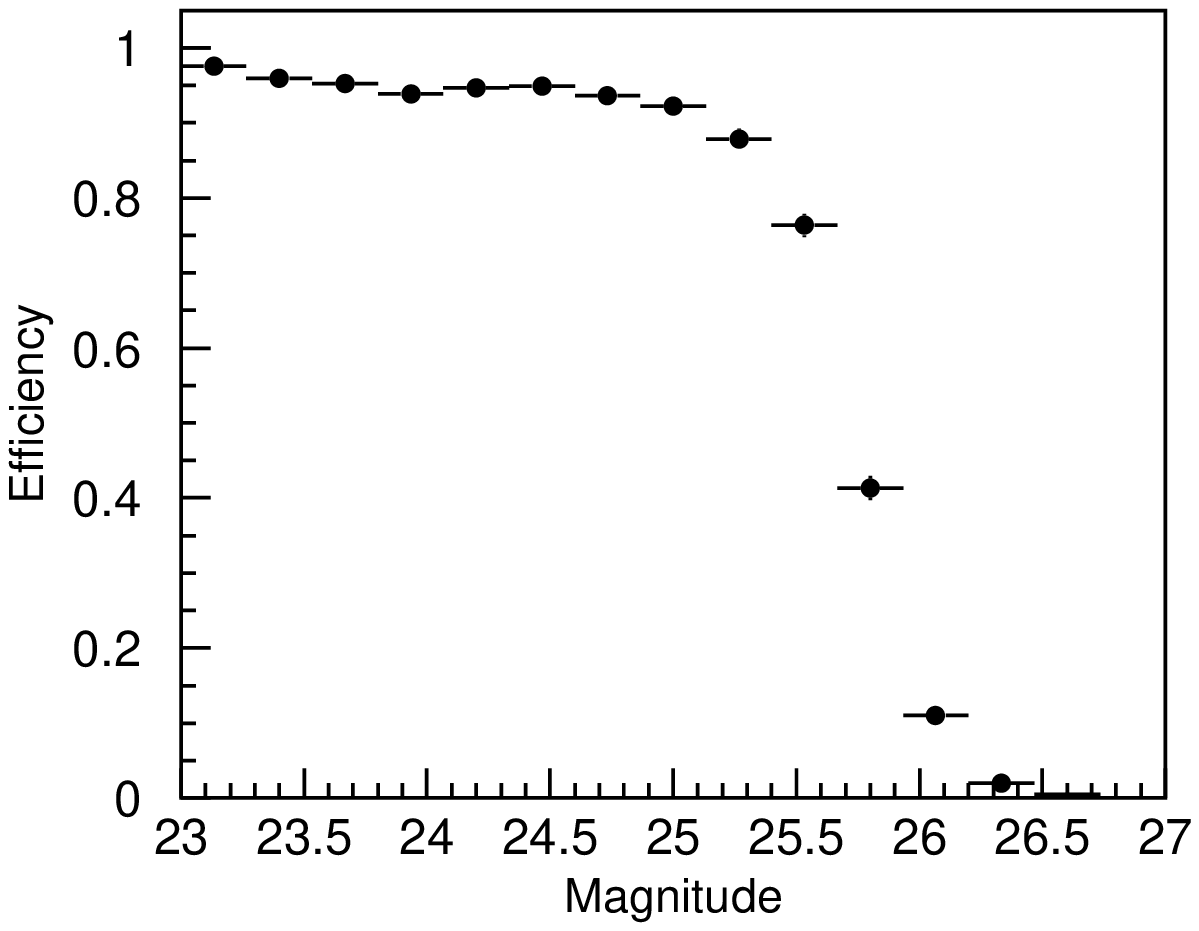} \\
\epsfxsize=3.0in\epsffile{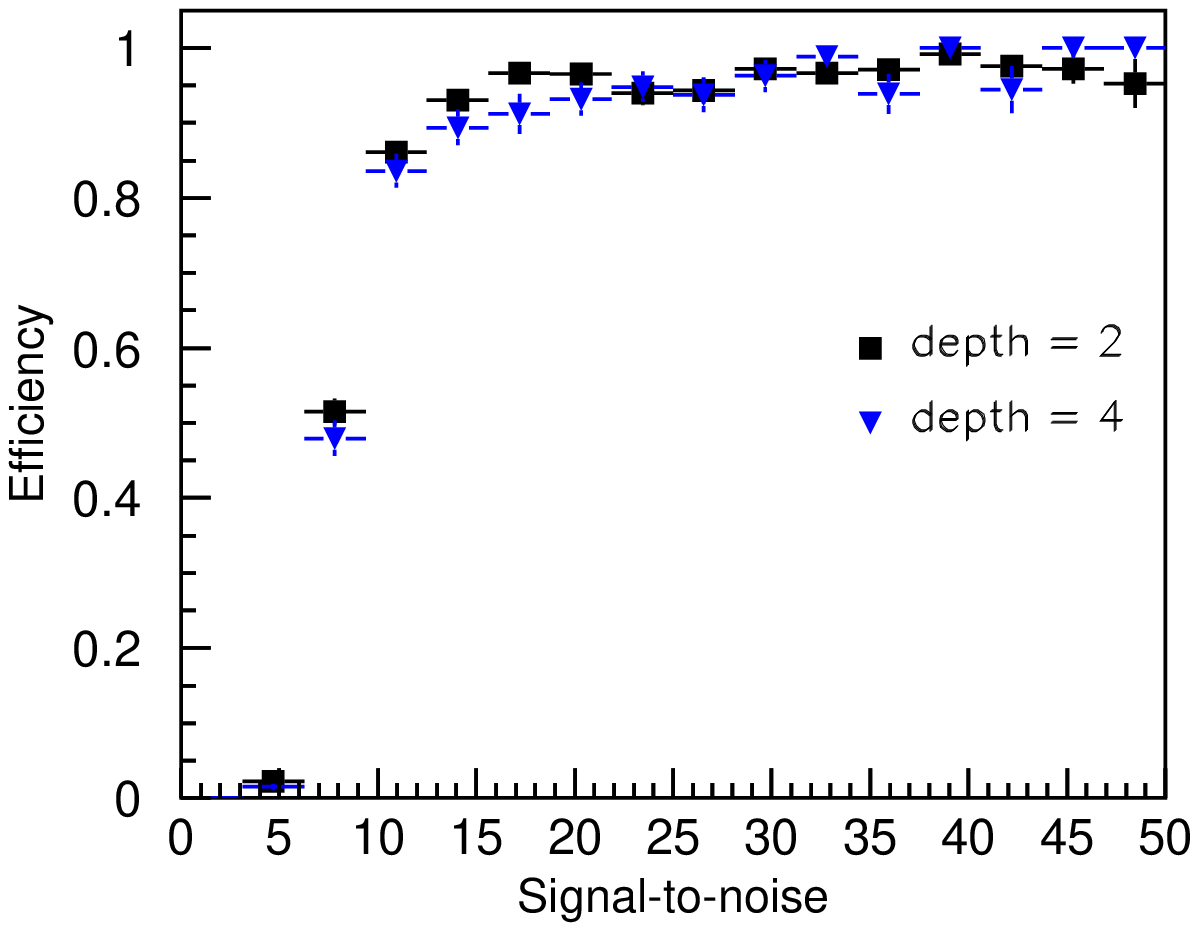}&
\epsfxsize=3.0in\epsffile{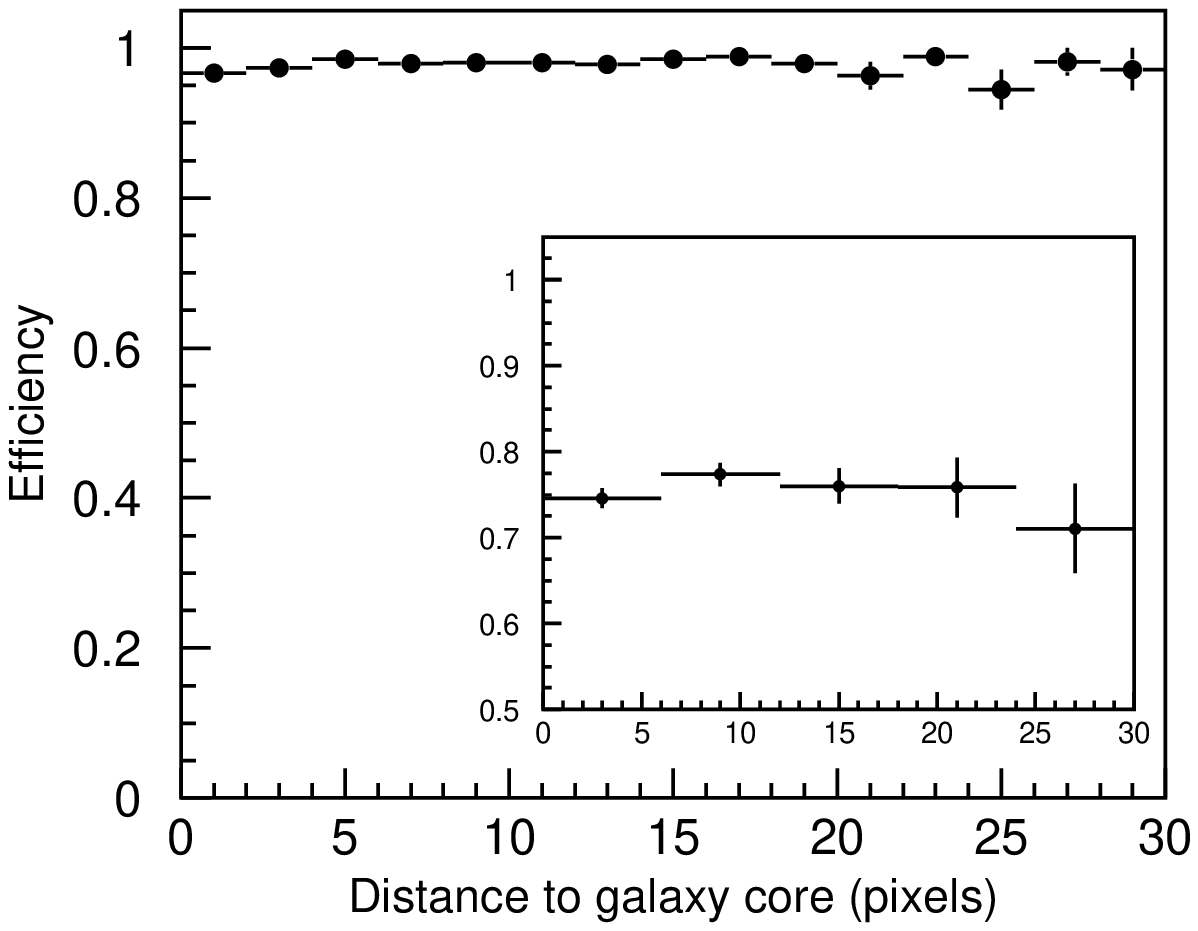} \\
\end{array}$
\end{center}
\caption{
The efficiency of the stage 1  supernova selection 
in the reference-subtracted search images.
The errors are assigned using binomial statistics.
The upper left plot shows the combined efficiency for all depths (see text for the definition of depth) 
of the reference image as a function of the candidate's SNR.
Overlaid as a solid line is the fit of the efficiency to the function in Eqn.~\ref{eqn:eff}.
The upper right plot shows the same as a function of the candidates' magnitude.
The lower left plot shows the efficiencies for two representative depths: 2 (squares) 
and 4 (down turned triangles).  The lower right plot 
shows the efficiency as a function of the supernova distance from the host galaxy core for 
all candidates with SNR $>$ 15; the insert shows the efficiency for the candidates
with SNR $\leq$ 15.
}
\label{fig:eff}
\end{figure}

We generated $\sim$13,000 fake supernovae (the 100 supernovae
used for the check of the manual scanning efficiency were a subset of this
sample).
The fake supernovae that pass the stage 1 selection cuts described
above are compared with the input list of fakes.
This allows us to calculate the efficiency of the selection cuts 
for the preliminary supernova selection.  This efficiency
is shown in Fig.~\ref{fig:eff} (upper left) as a function of the candidates' SNR,
and in Fig.~\ref{fig:eff} (upper right) as a function of the candidates' magnitude,
on the reference-subtracted search images.
Note that our reference images are not uniformly deep: they consist of 2, 4, or 5
combined epochs, depending on the tile of the GOODS field and the supernova's 
position on the tile (see Fig.~\ref{fig:goods}).    
Figure~\ref{fig:eff} (lower left) shows the supernova finding efficiency as a function
of the SNR for two representative cases: i) for all locations where
two epochs contribute to the reference data; and ii) for all locations
where there are four epochs that are available for the reference data.
We refer to these cases as ``depth 2'' and ``depth 4'', respectively.
It is evident that, within errors, for a given SNR, the efficiency 
is independent of the depth of the reference image at the location of the 
fake supernovae, as it should be.  We thus use the efficiency curve in Fig.~\ref{fig:eff} (upper left) 
that combines all of the depths, which we fit to the following four-parameter function:
\begin{equation}
\epsilon(SNR) = p_1 + \frac{p_2}{1+e^{p_3 \, (SNR - p_4)} }
\label{eqn:eff}
\end{equation}
where we obtain $p_1$ = 0.96, $p_2$ = $-$18.04, $p_3$ = 0.41, and $p_4$ = $-$1.34.
The resulting fit is also shown in Fig.~\ref{fig:eff} (upper left).

One concern in supernova searches is the potential loss of candidates
located close to the core of their host galaxies.  
Figure~\ref{fig:eff} (lower right) shows the efficiency as a 
function of the supernova's distance from the galaxy core.  It is apparent that the efficiency remains
essentially flat.

\subsection{Stage 2: The Multi-Epoch Selection}
\label{sec:multibandsearch}
The second stage of the supernova candidate selection is where we turn
to the multi-epoch photometric data in both filters.
Subtracting the stacked image of each epoch of the
search data from the reference data,
we calculate the candidates' SNRs in the subtracted data
and require that there be at least three epochs with a SNR $>$ 2, including at least 
two epochs with the SNR $>$ 3.  These cuts are designed to select
candidates with reasonably well-measured light curves.
Because the Bayesian technique described in~\ref{sec:typeia} provides
a powerful discrimination of Type Ia supernovae, these cuts can be very loose.
At the end of stage 2, we have 26 candidates in sample \#1, 
17 candidates in sample \#2, 9 candidates in sample \#3, and 5 candidates in sample \#4,
for a total of 57 candidates.   A list of these candidates is given in 
Tables~\ref{tab:57cands1},~\ref{tab:57cands2},~\ref{tab:57cands3}, 
and~\ref{tab:57cands4}, for samples \#1, \#2, \#3, and \#4, respectively.
The tables specify the supernova names and classifications (gold, silver, or bronze)
for the candidates that were also found in~\cite{bib:riess2} and in~\cite{bib:riess06}.  
The classification refers to 
the degree of belief in the typing of the candidate, with gold being certain.
For sample \#1, we have 8 gold and 1 silver Ia's, and 3 silver core-collapse (CC) supernovae.
For samples \#3 and \#4, we have 5 gold and 2 silver Ia's, and 1 gold and 1 silver CC.
There were 6 additional gold and silver Ia's found in~\cite{bib:riess2}
that failed our stage 2 cuts (SN-2003eu, SN-2002lg,  SN-2002fx, SN-2003ak, SN-2003eq, and SN-2003al)
because they did not have a sufficient number of epochs with high enough SNR.
In other words, these candidates fall below the threshold that is intentionally
set high enough that an automated Bayesian classification of candidates 
(discussed in section~\ref{sec:typeia}) may be possible.  Note also that
the failure of real SNe Ia to pass stage 2 cuts is taken into account in the 
control time calculation (Section~\ref{sec:ctandarea}).

Spectroscopic redshifts (of the host, the SN or both) were taken from
the following sources:~\cite{bib:strolger},~\cite{bib:riess2},
~\cite{bib:cohen},~\cite{bib:cowie},~\cite{bib:keck},~\cite{bib:lefevre},~\cite{bib:vanzella},
and~\cite{bib:riess06}.
In some cases, the spectroscopic redshift has been determined more than once. We find good 
agreement in such cases. If a spectroscopic redshift was not available,
we used photometric redshifts from~\cite{bib:wolf},~\cite{bib:strolger},
and~\cite{bib:private}.

The host galaxies of three candidates 
(candidates \#9 and \#25 in Table~\ref{tab:57cands1} and candidate \#12 in Table~\ref{tab:57cands2}) were observed with the Subaru Faint Object Camera and 
Spectrograph (FOCAS;~\cite{bib:kash}) on May 17, 2007
All three host galaxies
were observed with the 300R grism and the SO58 order sorting filter, 
resulting in spectra covering the 5800-10000\,\AA\ spectral region with
a resolving power of $\sim$ 300.  Single emission lines were detected in the first
two galaxies.  If these lines are due to the [OII] doublet at 3727 \AA, then 
the redshifts of these sources are $z$ = 1.143 $\pm$ 0.001 and $z$ = 0.618 $\pm$ 0.001,
respectively.  The first measurement confirms the redshift reported in~\cite{bib:strolger}.
The second measurement is new.  Although the continuum of the third galaxy was 
detected, no clear spectral features are apparent, so we used the photometric redshift instead.

The typical error in the redshift that is measured spectroscopically is 
0.001, if the redshift was determined from host galaxy lines, or 0.01, if 
the redshift was determined from supernova features.  For photometric
redshifts, the error is larger, ranging from 0.05 to as high as 0.4.
The source of the redshift errors is listed in the tables as well.
Precision photometric measurements for previously unpublished candidates will be
made available in~\cite{bib:suzuki}.
\begin{table}[htbp]
  \begin{center}
\tiny
    \begin{tabular}{|c|c|c|c|c|c|c|c|c|}
    \hline\hline
       Cand & RA (2000.) & Dec (2000.) & Redshift & Error & Source & Ref & Comment & $P({\rm Ia}|\{D_i\},z)$\\
     \hline\hline
     1 & 12:37:06.938 & +62:09:15.81 & 0.53 & 0.25    &   host phot  & M\&D  &    & 1.0  \\ 
     2 & 12:37:01.537 & +62:11:28.66 & 0.778 & 0.001  &   host spec  & C04   &    & 1.0 \\ 
     3 & 12:36:56.336 & +62:11:55.65 & 0.83 & 0.10    &   host phot  & M\&D  &    & 1.0  \\ 
     4 & 12:37:49.350 & +62:14:05.71 & 0.41 & 0.01    &   spec       & S04   &   silver CC (SN-2002kl) & 0.3  \\ 
     5 & 12:36:21.291 & +62:11:01.24 & 0.633 & 0.001  &   host spec  & TKRS  &    & 0.9  \\ 
     6 & 12:37:08.396 & +62:14:23.98 & 0.564 & 0.001  &   host spec  & TKRS  &    & 0.9  \\ 
     7 & 12:37:40.658 & +62:20:07.42 & 0.741 & 0.001  &   host spec  & TKRS  &    & 1.0  \\ 
     8 & 12:36:16.850 & +62:14:37.30 & 0.71  & 0.05   &   host phot  & S04   &  bronze Ia (SN-2002kh)  & 1.0  \\ 
     9 & 12:37:28.421 & +62:20:39.56 & 1.141 & 0.001  &  (host + SN) spec    & R04   & gold Ia (SN-2002ki)& 1.0  \\
     10 & 12:36:38.130 & +62:09:52.88 & 0.513 & 0.001 &  host spec   & S04, TKRS  & silver CC (SN-2003bc) & 0.0 \\ 
     11 & 12:37:25.126 & +62:13:16.98 & 0.67 & 0.01   &  SN spec     & R04   & gold Ia (SN-2003bd)        & 1.0  \\ 
     12 & 12:36:24.506 & +62:08:34.84 & 0.954 & 0.001 &  host spec & S04, TKRS   & silver CC (SN-2003bb)      & 0.8  \\
     13 & 12:36:27.828 & +62:11:24.71 & 0.66 & 0.05   &  host phot   & S04   & bronze CC (SN-2003ew)      & 1.0  \\ 
     14 & 12:37:19.723 & +62:18:37.23 & 1.27 & 0.01   & SN spec      & R04   & gold Ia (SN-2003az)        & 1.0  \\ 
     15 & 12:37:15.208 & +62:13:33.55 & 0.899 & 0.001 & (host + SN) spec     & R04, TKRS   & gold Ia (SN-2003eb) & 0.0  \\ 
     16 & 12:36:55.441 & +62:13:11.46 & 0.954 & 0.001 & (host + SN) spec     & R04, TKRS   & gold Ia (SN-2003es) & 1.0 \\ 
     17 & 12:36:33.179 & +62:13:47.34 & 0.54  & 0.05  &  host phot   & S04   & bronze Ia (SN-2003en)      & 0.9  \\ 
     18 & 12:36:57.900 & +62:17:23.24 & 0.529 & 0.001 &  host spec   & TKRS  &                         & 1.0 \\ 
     19 & 12:36:39.967 & +62:07:52.12 & 0.48  & 0.05  &  host phot   & S04   & bronze CC (SN-2003dz)      & 0.9  \\ 
     20 & 12:36:31.772 & +62:08:48.25 & 0.46  & 0.05  &  host phot   & S04   & bronze CC (SN-2003dx)      & 0.0  \\ 
     21 & 12:37:28.992 & +62:11:27.36 & 0.935 & 0.001 &  host spec   & S04, TKRS   & silver Ia (SN-2003lv) & N/A \\ 
     22 & 12:37:09.189 & +62:11:28.17 & 1.340 & 0.001 & (host + SN) spec  & R04, TKRS & gold Ia (SN-2003dy) & 1.0\\ 
     23 & 12:37:12.066 & +62:12:38.04 & 0.89  & 0.05  &  host phot  & S04    & bronze CC (SN-2003ea)       & 0.4 \\ 
     24 & 12:36:15.925 & +62:12:37.38 & 0.286 & 0.001 &  host spec  & S04, TKRS & bronze CC (SN-2003ba)    & N/A \\ 
     25 & 12:36:26.718 & +62:06:15.16 & 0.618 & 0.001 &  host spec  & this paper  &                     & N/A \\ 
     26 & 12:36:26.013 & +62:06:55.11 & 0.638 & 0.001 & (host + SN) spec   & R04, TKRS & gold Ia (SN-2003be) & 1.0 \\
     \hline\hline
    \end{tabular}
 \end{center}
\caption[]
{\label{tab:57cands1}
The candidates selected at the end of stage 2 for sample \#1.
Listed are the candidates' coordinates, redshifts, errors on the redshifts,
the sources used for the redshift and redshift error determination, the references
for the sources, 
and $P({\rm Ia}|\{D_i\},z)$ defined in section~\ref{sec:typeia}
(the ``N/A'' stands for a special category of candidates designated
as ``anomalies'', as described in section~\ref{sec:typeia}).
For the candidates found in~\cite{bib:riess2}, the tables also list the
supernovae' name and classification (gold, silver, or bronze).
C00 is~\cite{bib:cohen},
H03 is~\cite{bib:horsh},
CO4 is~\cite{bib:cowie},
VVDS is~\cite{bib:lefevre}, 
M\&D is~\cite{bib:private},
S04 is~\cite{bib:strolger}, 
R04 is~\cite{bib:riess2},
TKRS is~\cite{bib:keck},
W04 is~\cite{bib:wolf},
F2 is~\cite{bib:vanzella}, and 
R07 is~\cite{bib:riess06}.
}
\normalsize
\end{table}
\begin{table}[htbp]
 \tiny
  \begin{center}
    \begin{tabular}{|c|c|c|c|c|c|c|c|c|c|}
    \hline\hline
       Cand & RA (2000.) & Dec (2000.) & Redshift & Error & Source & Ref &  Comment & $P({\rm Ia}|\{D_i\},z)$\\
     \hline\hline
     1 & 12:36:20.889 & +62:10:19.24 & 1.10   & 0.28  & host phot & M\&D  &  & 1.0\\ 
     2 & 12:36:29.474 & +62:11:41.40 & 1.35   & 0.40  & host phot & M\&D  &  & 0.0 \\ 
     3 & 12:36:19.901 & +62:13:47.67 & 0.535  & 0.001 & host spec & TKRS  &  & 0.0 \\
     4 & 12:36:27.131 & +62:15:09.27 & 0.794  & 0.001 & host spec & TKRS  &  & 0.0 \\ 
     5 & 12:36:32.238 & +62:16:58.38 & 0.437  & 0.001 & host spec & TKRS  &  & 0.4\\ 
     6 & 12:38:03.689 & +62:17:12.23 & 0.280  & 0.001 & host spec & CO4   &  & 0.4 \\   
     7 & 12:37:09.495 & +62:22:15.37 & 1.61   & 0.34  & host phot & M\&D  &  & 1.0\\
     8 & 12:37:06.772 & +62:21:17.46 & 0.406  & 0.001 & host spec & TKRS  &  & 0.1 \\ 
     9 & 12:36:26.694 & +62:08:29.74 & 0.555  & 0.001 & host spec & TKRS  &  & 0.8 \\ 
     10 & 12:36:54.125 & +62:08:22.21 & 1.39  & 0.01  & SN spec   & R07   & gold Ia (HST04Sas)  & 1.0 \\ 
     11 & 12:36:34.363 & +62:12:12.55 & 0.457 & 0.001 & (host + SN) spec  & TKRS, R07 & gold Ia (JST04Yow) & 1.0 \\ 
     12 & 12:37:33.918 & +62:19:21.75 & 0.88  & 0.38  & host phot & M\&D &  & 1.0 \\ 
     13 & 12:36:34.853 & +62:15:48.86 & 0.855  & 0.001  & (host + SN) spec   & R07, TKRS & gold Ia (HST04Man)  & 1.0 \\ 
     14 & 12:36:36.009 & +62:17:31.97 & 0.60  & 0.15  &  host phot & M\&D &  & 0.2  \\
     15 & 12:36:55.214 & +62:13:03.75 & 0.952 & 0.004 &  (host + SN) spec & C00, R07 & gold Ia (HST04Tha)  & 1.0 \\ 
     16 & 12:37:48.435 & +62:13:34.85 & 0.839 & 0.001 &  host spec & TKRS &   & 1.0 \\
     17 & 12:36:01.542 & +62:15:55.16 & 0.086 & 0.001  &  host spec  & H03 &   &  N/A \\
     \hline\hline
    \end{tabular}
 \end{center}
\caption[]
{\label{tab:57cands2}
Same as Table~\ref{tab:57cands1} for sample \#2.   The redshift of candidate \#17 is uncertain,
as the possible host galaxy is 7$\arcsec$ away.  Leaving this redshift as unconstrained does
not change our results.
}
\normalsize
\end{table}
\begin{table}[htbp]
\tiny
  \begin{center}
  \begin{center}
    \begin{tabular}{|c|c|c|c|c|c|c|c|c|}
    \hline\hline
       Cand & RA (2000.) & Dec (2000.) & Redshift & Error & Source & Ref  & Comment& $P({\rm Ia}|\{D_i\},z)$\\
     \hline\hline

1 & 03:32:18.072 & -27:41:55.83 & 0.88  & 0.05   &  host phot   & S04  & silver Ia (SN-2002fy) & 0.9 \\ 
2 & 03:32:13.002 & -27:42:05.75 & 0.421 & 0.001  &  host spec   & VVDS &  & 0.0 \\ 
3 & 03:32:37.511 & -27:46:46.40 & 1.30  & 0.01   &  SN spect    & R04 &  gold Ia (SN-2002fw)  & 1.0\\ 
4 & 03:32:05.060 & -27:47:02.96 & 0.976 & 0.001  &  host spec   & VVDS &  & N/A \\
5 & 03:32:17.309 & -27:46:23.74 & 0.13  & 0.01   &  phot        & W04 & & 0.0 \\ 
6 & 03:32:48.598 & -27:54:17.14 & 0.841 & 0.001  & host spec    & S04, VVDS & silver CC (SN-2002fz)    & 0.9 \\ 
7 & 03:32:22.751 & -27:51:09.65 & Unconstrained & Unconstrained &  phot   & S04 & bronze CC (SN-2002fv)    & 0.0 \\
8 & 03:32:42.441 & -27:50:25.08 & 0.58 & 0.01 &  spec   & S04 & gold CC (SN-2002kb)      & N/A \\
9 & 03:32:38.082 & -27:53:48.15 & 0.987 & 0.001 &  host spec & S04, VVDS & bronze Ia (SN-2002ga)  & 1.0 \\ 
     \hline\hline
    \end{tabular}
 \end{center}
 \end{center}
\caption[]
{\label{tab:57cands3}
Same as Table~\ref{tab:57cands1} for sample \#3.  The redshift of candidate \#4 is uncertain,
as the possible host galaxy is 4$\arcsec$ away.  Leaving this redshift as unconstrained does
not change our results.
}
\normalsize
\end{table}
\begin{table}[htbp]
\tiny
  \begin{center}
    \begin{tabular}{|c|c|c|c|c|c|c|c|c|}
    \hline\hline
       Cand & RA (2000.) & Dec (2000.) & Redshift & Error & Source & Ref  & Comment & $P({\rm Ia}|\{D_i\},z)$\\
     \hline\hline
1 & 03:32:24.782 & -27:46:18.07 & 1.306 & 0.001 &  host spec         & R04, F2 &  gold Ia (SN-2002hp)   & 1.0\\ 
2 & 03:32:22.522 & -27:41:52.26 & 0.526 & 0.001 &  (host + SN) spec  & R04 & gold Ia (SN-2002hr) & 1.0\\ 
3 & 03:32:22.318 & -27:44:27.04 & 0.738 & 0.001 &  host spec         & R04, F2 & gold Ia (SN-2002kd)   & 1.0\\ 
4 & 03:32:05.382 & -27:44:29.76 & 0.91 & 0.05   &  host phot         & S04 & silver Ia (SN-2003al) & 1.0 \\ 
5 & 03:32:34.648 & -27:39:58.18 & 0.214 & 0.001 &  (host + SN) spec  & R04, VVDS & gold Ia (SN-2002kc)   & 1.0\\ 
     \hline\hline
    \end{tabular}
 \end{center}
\caption[]
{\label{tab:57cands4}
Same as Table~\ref{tab:57cands1} for sample \#4.
}
\normalsize
\end{table}

Note that the redshifts of candidate \#17 in Table~\ref{tab:57cands2} and
 candidate \#4 in Table~\ref{tab:57cands3} are uncertain, since the assumed host galaxies
of the supernova candidates are 7$\arcsec$ and 4$\arcsec$ away, respectively.  However, we have verified
that if we leave the redshifts of these candidates as unconstrained, it does not affect 
our final results.

\subsection{Stage 3: The Identification of Type Ia Supernovae}
\label{sec:typeia}
The candidates that have been selected in stages 1 and 2 are assumed to be 
real transient objects, most likely supernovae, and must now be classified by type.
With only scarce photometric data available, we turn to the Bayesian
method of classifying supernovae described in~\cite{bib:ourpaper}. 

Photometric typing of supernovae has been described 
in~\cite{pozn02},~\cite{riess2004},~\cite{john} and~\cite{sull1}, among others
Most of the existing methods rely on color-color or color-magnitude 
diagrams for supernova classification.

In our method, we consider five possible supernova 
types (``normal'' Ia~\citep{bib:branch}, Ibc, IIL, IIP, and IIn).  We make use of the best
currently available supernova multicolor lightcurve templates for each type.  When improved
supernova templates are available, they can be easily worked into the method.
We calculate the probability  that a given supernova candidate with photometric data $\{D_i\}$,
where $i$ is the index for the number of observational epochs, and redshift $z$ is 
a Type Ia supernova.  By virtue of the Bayes theorem, this probability is given by:
\begin{equation}
P({\rm Ia}|\{D_i\},\,z)= \frac{ \int_{\vec{\theta}} P(\{D_i\},z|\vec{\theta},{\rm Ia})   P(\vec{\theta},{\rm Ia}) d\vec{\theta}} {\sum_{T} \int_{\vec{\theta}} \, P(\{D_i\},\,z|\vec{\theta},T)P(\vec{\theta},T) d\vec{\theta}}.
\label{eqn:bayes}
\end{equation}
where $z$ is the measured supernova redshift; $\vec{\theta}$ are the parameters that characterize
a given supernova type; $\{D_i\}$ are the data in both F850LP and F775W; 
$P(\{D_i\}\,z| \vec{\theta}, T)$ is the probability density to obtain  data $\{D_i\}$ and redshift $z$
for supernova type $T$; $P(\vec{\theta},T)$ contains prior information about type $T$ supernovae;
and the denominator contains the normalization (the sum) over all five supernova types $T$ considered.
The parameters $\vec{\theta} \equiv (\bar{z}, t_{\rm diff},s,M,R_V,A_V)$
are: $\bar{z}$ is the true supernova redshift; $t_{\rm diff}$ is the time difference between 
the dates of maximum light for the template and the data;
$s$ is the stretch parameter~\citep{bib:perl}, which parametrizes the width of the 
light curve (if $T$ = Ia);  $M$ is the absolute  magnitude in the restframe $B$-band at maximum light;
and $A_V$ and $R_V$ are the Cardelli-Clayton-Mathis 
interstellar extinction parameters~\citep{bib:ccm}.
We marginalize (integrate over) these parameters as described below.

Suppose that we have a photometric template, $\{\bar{D}(\vec{\theta},T)_i\}$, for 
the expected light curve for a supernova of type $T$ at a given redshift, $\bar{z}$.
In this work, we use the templates from P. E. Nugent\footnote[3]{See http://supernova.lbl.gov/$\sim$nugent/nugent\_templates.html}, 
which 
extend both into the UV (below 3460 \AA\, in the supernova
rest frame) and into far red and IR (above 6600 \AA\, in the supernova rest frame) regions.
Is it assumed that the measured light curve flux, $\{D_i\}$, can fluctuate from the template $\{\bar{D}(\vec{\theta},T)_i\}$ 
according to Gaussian statistics.  It is also assumed that 
the probability of measuring redshift $z$ fluctuates
around a mean $\bar{z}$ according to Gaussian statistics as well.  Therefore,
\begin{equation}
P(\{D_i\},z |\vec{\theta}, T)  = \frac{{\rm exp}(-\frac{(z-\bar{z})^2}{2\delta z^2})} {\sqrt{2\pi} \delta z}\prod_{i=1}^{n_{epochs}} \frac{{\rm exp}({-\frac{(\bar{D}(\vec{\theta},T)_i-D_i)^2}{2 \delta D_i^2}}) }{\sqrt{2\pi} \delta D_i },
\label{equation:likelihood}
\end{equation}
where $\delta D_i$ are photometric measurement errors for epoch $i$, 
and $\delta z$ is the measurement error for the redshift $z$.
Note that we assume no errors on the supernova templates themselves; we take them to represent the best currently available
knowledge of the supernova behavior.  However, it is also worth noting that 
various parameters that characterize a given template (\emph{e.g.}, the peak
restframe $B$-band magnitude, the stretch parameter for Ia's, \emph{etc.}) are varied
as described below, thus effectively representing some template variations.

The prior $P(\vec{\theta}, T)$ contains all the available information about the 
behavior of type $T$ supernovae, expressed in terms of parameters ${\vec{\theta}}$.
We assume that all constituents of  ${\vec{\theta}}$ can be divided as 
follows where $t_{\rm diff}$, $\bar{z}$ and $T$, $M$, $R_V$ and $A_V$ and $s$ are independent.
\begin{equation}
P(\vec{\theta} ,T ) =  P(t_{\rm diff}|\bar{z},T) \, P(M|\bar{z},T)\, P(s|\bar{z},T)\, P(R_V,A_V|\bar{z},T)\, P(\bar{z},T).
\label{equation:prior}
\end{equation}
The assumed independence of the parameters is certainly an 
oversimplification.  For example, one would expect the stretch and magnitude parameters to be
correlated (although the true values of these two parameters should be independent of
$t_{\rm diff}$, $R_V$ and $A_V$).  Ignoring the correlation
might conceivably lead to an overestimation of the probabilities
for very bright Type Ia's with a small stretch parameter, or 
very dim Type Ia's with a large stretch parameter.  However, 
we are exploring every possible combination of stretch and magnitude
parameters; the ``correct'' combination should naturally be
a better ``fit'' to the data, thus acquiring a larger weight 
than all the other ones.

\begin{figure}[!htb]
 \begin{center}
    \includegraphics[height=4.0in, width=5.0in]{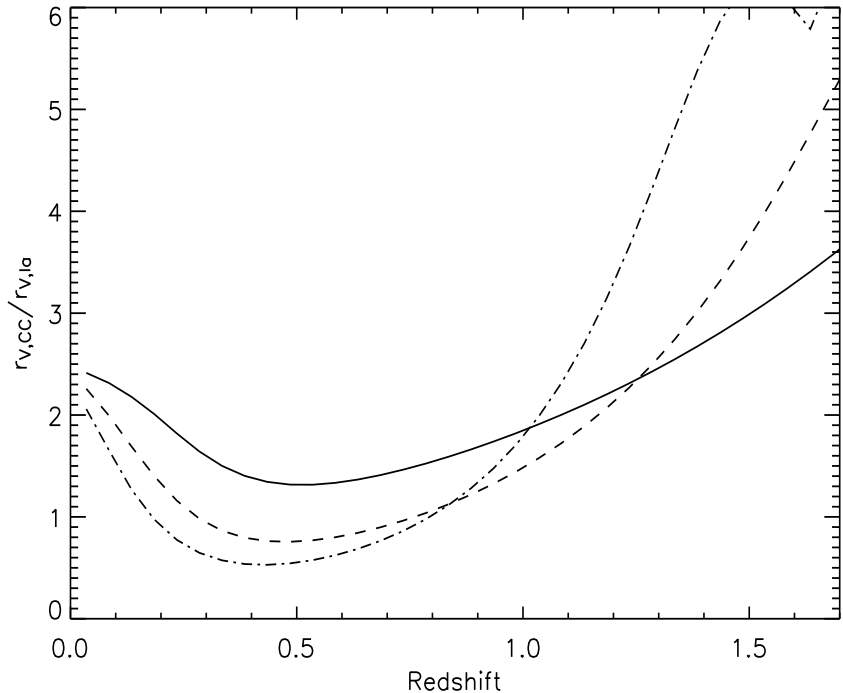}
  \end{center}
\caption[]
{\label{fig:ratio_models}
The three models for the ratio of the CC/Ia rates as a function of redshift based on~\cite{bib:dahlen2}.
The solid line is for the time delay parameter $\tau$ = 1 Gyr; the dashed line is 
for $\tau$ = 2 Gyr; and the dashed-dotted line is for $\tau$ = 3 Gyr.
}
\end{figure}

The prior $P(\bar{z},T)$ includes the relative rates of the various supernova
types as a function of redshift.  Unfortunately, these rates are not well known, especially at high redshift.  
We will thus consider three different models for the ratio of the CC supernova rates
to the Ia supernova rates.  The models are based on~\cite{bib:dahlen2}, and
shown in Fig.~\ref{fig:ratio_models}.  They correspond to three
different values of the characteristic time delay parameter $\tau$:
$\tau$ = 1 Gyr, 2 Gyr, and 3 Gyr.
Based on~\cite{bib:dahlen2}, we will also assume
that the relative (rounded-off) fractions of the CC supernovae are
$f_{Ibc}$ = 0.27, $f_{IIL}$ = 0.35, $f_{IIp}$ = 0.35, and $f_{IIn}$ = 0.02,
for all three models, regardless of the redshift.

Note that the usage of these models does not bias our answer in any way, as we are not making 
any assumptions about  the \emph{absolute} rates of supernovae, but only about their relative rates.
If we assume all three models to be equally likely, then
the probability density $P(T)$ of observing a supernova of type $T$ for assumption $n$
about the relative rates of the CC to Ia supernovae is given by:
\begin{equation}
P(\bar{z},T) = \frac{R_n(\bar{z},T)}{\sum_{l=1}^{N_{models}} R_l(\bar{z},T)}
\end{equation}
where $R_n(\bar{z},T)$ is the rate of type $T$ supernovae
for model $n$, and $N_{models}$ = 3.

The difference in the dates of maximum light between the template and the data, 
 $t_{\rm diff}$, can also take on any value, making the prior 
$P(t_{\rm diff}|T)$ flat.  In practice, we shift the relative dates of maximum between 
the measured and the template light curves by increments of one day.
The marginalization of this parameter thus amounts to a sum
over a finite number (which we take to be 160) of such shifts.
\begin{equation}
P(t_{\rm diff}|\bar{z},T) = \frac{1}{{t^{max}_{\rm diff}-t^{min}_{\rm diff}}},
\end{equation}  
where the maximum $t^{max}_{\rm diff}$ and minimum $t^{min}_{\rm diff}$ set the limits on $t_{\rm diff}$.

The priors on $P(M|\bar{z}, T)$ and $P(s|\bar{z}, {\rm Ia})$ are taken to be Gaussian:
\begin{equation}
P(M|\bar{z},T) =\frac{e^{-\frac{(M-\bar{M})^2}{2 \delta M^2}}}{\sqrt{2 \pi} \delta M}.
\label{eqn:magrioria}
\end{equation}  
\begin{equation}
P(s|\bar{z},{\rm Ia}) =\frac{e^{-\frac{(s-\bar{s})^2}{2 \delta s^2}}}{\sqrt{2 \pi} \delta s}
\label{eqn:sprioria}
\end{equation}
A table of the mean magnitudes $\bar{M}$ and the standard deviations $\delta M$,
as well as the values for the mean stretch $\bar{s}$  and the standard deviation $\delta s$,
are given in~\cite{bib:ourpaper}.  For reference, we extract the mean magnitudes 
$\bar{M}$ in the restframe $B$-band from 
P. E. Nugent\footnote[4]{See http://supernova.lbl.gov/$\sim$nugent/nugent\_templates.html}, 
and the standard deviations, $\delta M$, from~\cite{bib:rich}. 
The stretch parameters are extracted from~\cite{bib:sull}.
Note that for non-Ia's, a complete set of ``virtual'' values for the stretch parameters
are inserted into Eqn.~\ref{eqn:sprioria} and then marginalized with a flat
prior (see Appendix B in~\cite{bib:ourpaper}).

The effects of interstellar extinction are difficult to parametrize
due to lack of generally accepted models for the behavior of the 
Cardelli-Clayton-Mathis parameters $A_V$ and $R_V$.  We compromise
by considering a case of no extinction and 
two cases of extinction with a moderate value of $A_V$ = 0.4 and two 
different values of $R_V$, 2.1 and 3.1.  The mathematical framework used
in the analysis easily allows for the implementation of real distributions
for $A_V$ and $R_V$, once they become standardized.
It is known that in simulations $A_V$ is sharply peaked near 0 (\emph{e.g.},~\cite{bib:hatano};~\cite{bib:riel}); therefore, not considering 
very large values of $A_V$ is reasonable.  
All three cases ($N_V = 3$) are considered
equally possible. In other words, we take:
\begin{eqnarray}
P(R_V,A_V|\bar{z},T) = \frac{1}{N_V}
\end{eqnarray}
It is certainly a simplified extinction model; however, it appears to be sufficient
as demonstrated by the largely successful typing of known Type Ia candidates
in two such diverse samples as the 73 SNLS-identified~\citep{bib:snls}
Type Ia's and the gold and silver Ia's in the HST GOODS data~\citep{bib:ourpaper}.  The method correctly
identified 69 out of the 73 SNLS Type Ia's.  For the remaining four candidates,
at least one filter band included wavelengths outside of the well-understood 
optical range in the supernova rest frame.
It also correctly identified 7 out of 8 gold and silver Ia's,
and 5 out of 5 gold and silver CC's.  Another consideration 
to note here is that extinction primarily affects the measured magnitudes, and 
our model already takes into account wide variations in the magnitudes
(Eqn.~\ref{eqn:magrioria}).

Putting everything together, we see that the numerator of Eqn.~\ref{eqn:bayes} is given by:
\begin{small}
\begin{eqnarray}
\label{eqn:final_Ia}
& & \int_{\vec{\theta}} \, P(\{D_i\},z| \vec{\theta}, {\rm Ia})  P(\vec{\theta}, {\rm Ia}) d{\vec{\theta}}  =  \nonumber \\
& & \sum_{\bar{z}=z_{min}}^{z_{max}} \frac{\Delta \bar{z}}{\sqrt{2 \pi} \delta z} \, e^{-\frac{(z-\bar{z})^2}{2 \delta z^2}} \, \frac{1}{\sum_{l=1}^{N_{models}} R_l(\bar{z},T)}\sum_{n=1}^{N_{models}} R_n(\bar{z},{\rm Ia}) \, \frac{1}{N_v}\sum_{n_v=1}^{N_V}\Delta N_v\sum_{M=M_{min}}^{M_{max}} \frac{\Delta M}{\sqrt{2 \pi} \delta M} \, e^{-\frac{(M-\bar{M})^2}{2 \delta M^2}} \,\frac{\Delta t_{\rm diff}} {{t^{max}_{\rm diff}-t^{min}_{\rm diff}}} \nonumber \\
& &  \sum_{s=s_{min}}^{s_{max}} \frac{\Delta s}{\sqrt{2 \pi} \delta s} e^{-\frac{(s-\bar{s})^2}{2 \delta s^2}} \, \prod_{i=1}^{n_{epochs}} \frac{{\rm exp}({-\frac{(\bar{D}_{j}-D_i)^2}{2\delta D_{i}^2}})}{\sqrt{2 \pi} \delta D_i}
\end{eqnarray}
\end{small}
for Ia's, and for types $T'$ that are non-Ia's, it is:
\begin{small}
\begin{eqnarray}
\label{eqn:final_nonIa}
& & \int_{\vec{\theta}} \, P(\{D_i\},z| \vec{\theta}, T')  P(\vec{\theta}, T') d{\vec{\theta}}  = \nonumber \\
& & \sum_{\bar{z}=z_{min}}^{z_{max}} \frac{\Delta \bar{z}}{\sqrt{2 \pi} \delta z} \, e^{-\frac{(z-\bar{z})^2}{2 \delta z^2}}\,
\frac{1}{\sum_{l=1}^{N_{models}} R_l(\bar{z},T)}\sum_{n=1}^{N_{models}} R_n(\bar{z},T') \, \frac{1}{N_v}\sum_{n_v=1}^{N_v}\Delta N_v\sum_{M=M_{min}}^{M_{max}} \frac{\Delta M}{\sqrt{2 \pi} \delta M} \, e^{-\frac{(M-\bar{M})^2}{2 \delta M^2}} \,\frac{\Delta t_{\rm diff}} {{t^{max}_{\rm diff}-t^{min}_{\rm diff}}} \nonumber \\
& & \prod_{i=1}^{n_{epochs}} \frac{{\rm exp}({-\frac{(\bar{D}_{j}-D_i)^2}{2\delta D_{i}^2}})}{\sqrt{2 \pi} \delta D_i}.
\end{eqnarray}
\end{small}
In Eqns.~\ref{eqn:final_Ia} and~\ref{eqn:final_nonIa}, we marginalize over parameters $\vec{\theta}$,
approximating the integration by summation.  
The range of redshifts [$z_{zmin}$, $z_{max}$] is taken to  be from 0 to 1.7 in the 
denominator of Eqn.~\ref{eqn:bayes}, and over a bin of interest in the numerator
(this point will be explained in more detail later in this section), and we take $\Delta z$ = 0.05.
The mean values of $z$ and the error on the $z$, $\delta z$, are given in 
Tables~\ref{tab:57cands1},~\ref{tab:57cands2},~\ref{tab:57cands3}, and~\ref{tab:57cands4}
for the candidates used in the analysis.
$\Delta t_{\rm diff}$ is one day, and $\Delta N_v$ = 1.
We sum $M$ from $M_{min}$ = -3$\delta M$ to $M_{max}$ = +3$\delta M$ with a total of 12 steps,
and we sum $s$ from $s_{min}$ = 0.65 to $s_{max}$ = 1.3  in 14 steps.  
For non-Type Ia's, a complete set of ``virtual'' values for the stretch parameters
are inserted into Eqn.~\ref{eqn:sprioria} and then marginalized with a flat
prior (see Appendix B in~\cite{bib:ourpaper}).

The probability that $\alpha^{\rm th}$ candidate is a Type Ia supernova
belonging to the $j^{\rm th}$ redshift bin,
[$\bar{z}_{j\, lower}$, $\bar{z}_{j \, upper}$], is thus:
\begin{equation}
P_j^\alpha = \frac{ \int_{\bar{z}_{j\, lower}}^{\bar{z}_{j\, upper}}d\bar{z} \, \int_{\vec{\theta}}  P(\{D_i\}, z| \vec{\theta}, {\rm Ia})  P(\vec{\theta}, {\rm Ia}) \,d\vec{\theta}} {\sum_{T}\int_{0}^{\infty}d\bar{z} \int_{\vec{\theta}} P(\{D_i\},z| \vec{\theta},T)  P(\vec{\theta},T)\,d\vec{\theta}}.
\end{equation}.

Let us now introduce the following variables:
\begin{itemize}
\item
 $N_j$ is the total count of the candidates contributing to the $j^{\rm th}$ redshift bin.
\item
 $P_j^{\alpha}$ is the Bayesian probability for each candidate $\alpha$  in the $j^{\rm th}$ redshift bin ($\alpha = [1,...,N_j]$).
\item
 $\{P^{\alpha}\}_j$ is the full set of probabilities for the candidates in the $j^{th}$ redshift bin.
\item
 $d_j$ is the most likely number of Ia candidates in the $j^{\rm th}$ redshift bin.
\end{itemize}
Our goal is to find $d_j$, as well as  the error on this number, given  $N_j$ and $P_j^{\alpha}$'s.

If $N_j$ is large, say of order 100 (which is the case for our Monte Carlo samples), then
$d_j$ can be simply evaluated as:
\begin{equation}
d_j = \sum_{\alpha = 1}^{N_j} P_j^\alpha,
\label{eqn:numcands}
\end{equation}
where the uncertainty on $d_j$ is given by the square root of the binomial and Poisson variances:
\begin{equation}
\Delta d_j = \sqrt{ \sum_{\alpha = 1}^{N_j} P_j^\alpha (1 - P_{j}^{\alpha}) + \sum_{\alpha = 1}^{N_j} P_j^\alpha}.
\label{eqn:numcandserr}
\end{equation}
Note that if all of the probabilities $P_j^\alpha$
were 1 (\emph{i.e.}, the candidates were all known to be Type Ia supernovae), using
Eqn.~\ref{eqn:numcands} would amount to a simple counting of the number of candidates,
and Eqn.~\ref{eqn:numcandserr} would become the usual $\sqrt{N_j}$ error for a large 
number of events $N_j$.

For a small number of events, $N_j$ $<$ 10, which is typically the case for our data
samples, using Eqn.~\ref{eqn:numcands} and~\ref{eqn:numcandserr} would be incorrect.
A more sophisticated approach is needed.
Let us define a variable $x_\alpha$ such that $x_\alpha = 1$
if the $\alpha^{\rm th}$ candidate is indeed a Type Ia and $x_\alpha = 0$
if it is not, so that there are  $k_j$ $\equiv$ $\sum_{\alpha=1}^{N_j} x_\alpha$
Type Ia's in this bin.  The probability to obtain $d_j$ is given by:
\begin{eqnarray}
P(d_j|\{P^\alpha \}_j) = \sum_{\{x_\alpha \}} P(d_j|\{x_\alpha \})P(\{x_\alpha \}|\{P^\alpha \}_j) = \\ 
\sum_{\{x_\alpha \}} \frac{P(\{x_\alpha \}|d_j)P(d_j)}{\int_{d_{j}=0}^{\infty} P(\{x_\alpha \}|d_j)P(d_j) d(d_j) }P(\{x_\alpha \}|\{P^\alpha \}_j),
\label{eqn:bayes2}
\end{eqnarray}
where the sum on $d_j$ can, in principle, extend to arbitrarily large values
(for example, if $N_j$ = 2, there is still a small but non-zero probability that 
$d_j$ can be 100).  We will assume a flat prior for $P(d_j)$, in which case the 
denominator integrates to unity.

The first term in Eqn.~\ref{eqn:bayes2} is a normalized Poisson distribution 
for the expected $d_j$ number of events while $k_j=\sum_{\alpha = 1}^{N_j}x_\alpha$ events are assumed to be in the $j^{th}$ bin:
\begin{equation}
P(\{x_\alpha \}|d_j) = \frac{d_j^{k_j}e^{-d_j}}{k_j!}~{\rm,\, where}~k_j=\sum_{\alpha =1}^{N_j}x_\alpha .
\end{equation}
The term $P(\{x_\alpha\}|\{P^\alpha \}_j)$ in Eqn.~\ref{eqn:bayes2} is the probability that
certain supernovae do or do not occupy the $j^{th}$ bin.  This probability 
is simply:
\begin{equation}
P(\{x_\alpha\}|\{P^\alpha \}_j) = \prod_{\alpha=1}^{N_j} [ P_j^\alpha x_\alpha + (1-P_j^\alpha)(1-x_\alpha)].
\end{equation}

Because we have no way of knowing \emph{a priori} which candidate belongs in the 
$j^{th}$ bin, we must sum over all possible $\{x_{\alpha}\}$'s:
\begin{equation}
P(d_j|\{P^\alpha\}_j) = \sum_{\{x_\alpha\}}\frac{d^{k_j}e^{-d_j}}{k_j!}\prod_{\alpha=1}^{N_j} [ P_j^\alpha x_\alpha + (1-P_j^\alpha)(1-x_\alpha)].
\label{eqn:thed}
\end{equation}
To obtain the best estimate for $d_j$, we must maximize $P(d_j|\{P^\alpha\} _j)$
given in Eqn.~\ref{eqn:thed}.  
In practice, this is done numerically for a range of test $d_j$'s from 0 to some maximum $d_{j\, max}$ (we arbitrarily take it 
to be 50) to  find out which $d_j$ maximizes the probability.

Let us consider an example.  Suppose that we have two supernovae in a given bin, 
with probabilities of being Ia's given by $P^1$ = 0.8 and $P^2$ = 0.9.  The possible permutations
of $x_\alpha$'s would be (0,0), meaning that neither candidate is a Type Ia; (0,1) and (1,0), meaning
that only one candidate is a Type Ia; and (1,1), meaning that both candidates are Ia's.
Then we need to maximize
\begin{equation}
\frac{d^0 e^{-d}}{0!}  (1 - 0.8)  (1 - 0.9) + \frac{d^1 e^{-d}}{1!}  0.8  (1 - 0.9) + \frac{d^1 e^{-d}}{1!}  ( 1 - 0.8)  0.9 + \frac{d^2 e^{-d}}{2!}  0.8 \times 0.9
\end{equation}
as a function of $d$.  For this particular example, the best estimate for the number of Type Ia's is in fact $1.68^{+2.62}_{-0.58}$, where the errors are estimated as described below.

To evaluate the uncertainty on $d_j$, we 
find the 68\% confidence regions for $d_j$, [$d_j-\sigma_{j\,low}$, $
d+\sigma_{j\,high}$], by solving:
\begin{equation}
16\% = \int_0^{d_j-\sigma_{j\,low}} P(d_j|\{P^\alpha\}_j)d(d_j) = \int_{d+\sigma_{j\,high}}^\infty P(d_j|\{P^\alpha\}_j)d(d_j)
\label{eqn:error}
\end{equation}
In the case where $d_j<<1$, we set $\sigma_{j\,low} = 0$ and find
$\sigma_{j\,high}$ by satisfying:
\begin{eqnarray}
32\%=\int_{\sigma_{j,high}}^\infty P(d_j|\{P^\alpha\}_j)d(d_j).
\end{eqnarray} 

We assume that all candidates whose redshift is within $\pm$3 $\delta z$ 
of the $j^{th}$ bin's boundaries 
(where $\delta z$ is the uncertainty on the candidates' redshift, listed in 
Tables~\ref{tab:57cands1},~\ref{tab:57cands2},~\ref{tab:57cands3}, and~\ref{tab:57cands4})
will contribute to this bin.
Note that in this formulation, a single candidate with a poorly known redshift may 
have a probability distribution that spans several redshift bins.

We calculate $P({\rm Ia} |\{D_i\},z)$ for all 57 candidates.  If a given 
candidate's $P(\{D_i\},z | \vec{\theta},T) \, P(\vec{\theta},T)$
is less than 10$^{\rm -15}$ for all types $T$ ,
it is considered to be an ``anomaly'' and is excluded from further consideration.
The 10$^{\rm -15}$ cut was chosen because it is much smaller than 
the values calculated for simulated supernovae in the Monte Carlo.
This method thus excludes any need for the often subjective
and time-consuming decision on whether or not a candidate might be a 
supernova of a given type;
all dubious candidates are weighted appropriately and
left in the sample for the probability to decide.

It is a good sanity check to examine the values of 
$P({\rm Ia} |\{D_i\},z)$ for the gold and silver Ia candidates
from~\cite{bib:riess2}.  
Tables~\ref{tab:57cands1},~\ref{tab:57cands2},~\ref{tab:57cands3}, and~\ref{tab:57cands4}
list $P({\rm Ia} |\{D_i\},z)$ (with [$\bar{z}_{j\, lower}$, $\bar{z}_{j \, upper}$] = [0.0, 1.7])
for all of the candidates.  Several candidates have ``N/A'' listed for 
$P({\rm Ia} |\{D_i\},z)$: these are the ``anomalous'' candidates, as described above.
It is apparent that the gold and silver Ia candidates are among the largest contributors
to a given redshift bin.  All but one of them, SN-2003eb, have probabilities $\geq$ 0.8.
SN-2003eb has only two epochs (epochs 4 and 5 of the GOODS dataset) with 
``appreciable'' SNR ($>$ 10) in both F775W and F850LP bands.  
One silver Ia candidate, SN-2003lv, appears to have a rare residual cosmic ray contamination
in the F775W band,
making it appear inconsistent with any of the supernova types considered.
Three silver core-collapse supernovae, SN-2002kl, SN-2003bb, and SN-2002fz
have the probabilities of being Ia's of 0.3, 0.8, and 0.9 respectively.
They are in fact most consistent with being IIn's; however, because the fraction 
of IIn's is heavily de-weighted among CC supernovae ($f_{IIn}$ = 0.02), their resulting 
$P({\rm Ia} |\{D_i\},z)$ are higher than one would have expected.
How  much do our assumptions about the fractions of various supernova  
types among the CC supernovae
influence our answer? As we will see in Section~\ref{sec:priors}, if we assume  
that all CC types are equally
likely and that the ratio of the CC to Ia rates is redshift  
independent, the changes  to our final results are within the quoted  
uncertainties.

Another sanity check is to make sure that the candidates with 
low $P({\rm Ia} |\{D_i\},z)$'s are not all of a particular class (\emph{e.g.}, Ibc's).
We have verified that indeed they are not.

It is worth noting that variable objects other than supernovae, such as AGNs, 
are selected during the first selection stage. If some
of these objects also pass the second selection stage, they are 
unlikely to bias the results significantly, as the
specifically designed cuts in the third stage would likely 
reject such candidates.  As an extra check, we verified that none of the candidates listed in
Tables~\ref{tab:57cands1},~\ref{tab:57cands2},~\ref{tab:57cands3}, and~\ref{tab:57cands4}
that are close (within 3 pixels) to the core of their host galaxies have a matching
x-ray-bright object in the Chandra Deep Field catalogs~\citep{bib:chandra1, bib:chandra2}.
The only questionable candidate that might have a matching object is candidate \#3
in Table~\ref{tab:57cands2}; however, its $P({\rm Ia} |\{D_i\},z)$ never
exceeds $\sim$10$^{-6}$ for any redshift bin considered.

In order to estimate $d_j$'s, one must select some kind of redshift binning.
One must be careful about the selection of the redshift bins in an analysis
whose goal is to estimate the supernova rates, because 
the use of binning averages the behavior of the rates over the width of the bin.
However, the uncertainty in the candidates'
redshifts forces us to use finite bins -- or, in other words, it does not make
sense to use infinitely narrow bins when there is significant uncertainty
in the candidate redshifts.
For our analysis, we choose the width of the bins to be $\Delta \bar{z}$ = 0.1.
Table~\ref{tab:obscount} lists the numbers of observed candidates in these bins,
as well as their uncertainties, for the four samples listed in Table~\ref{tab:samples}
($d_j^m$ refers to a number of candidates in the $j^{\rm th}$ redshift bin
for the $m^{\rm th}$ sample).
All the uncertainties reflect a 68\% confidence region.
In order to calculate the total numbers of supernovae, $d_j$,
we use the procedure described above on the combined candidates from all four samples.
In other words, the total $d_j$ is not a trivial sum of the probability distributions 
of the $d_j^m$'s.
\begin{table}[htbp]
\small
  \begin{center}
    \begin{tabular}{|c|c|c|c|c|c|}
    \hline\hline
       Redshift bin             & $d_j^1$ & $d_j^2$ & $d_j^3$ & $d_j^4$ & Total\\
     \hline\hline
0.0  $\leq$  $z$ $<$ 0.1 &  0.00$^{+1.13}_{-0.00}$ & 0.00$^{+1.13}_{-0.00}$ & 0.00$^{+1.13}_{-0.00}$  & 0.00$^{+1.13}_{-0.00}$ & 0.00$^{+1.13}_{-0.00}$\\
0.1  $\leq$  $z$ $<$ 0.2 &  0.00$^{+1.13}_{-0.00}$ & 0.00$^{+1.13}_{-0.00}$ & 0.00$^{+1.13}_{-0.00}$  & 0.00$^{+1.13}_{-0.00}$ & 0.00$^{+1.13}_{-0.00}$\\
0.2  $\leq$  $z$ $<$ 0.3 &  0.00$^{+1.13}_{-0.00}$ & 0.00$^{+1.13}_{-0.00}$ & 0.00$^{+1.13}_{-0.00}$  & 0.00$^{+1.13}_{-0.00}$ & 0.00$^{+1.15}_{-0.00}$\\
0.3  $\leq$  $z$ $<$ 0.4 &  0.00$^{+1.17}_{-0.00}$ & 0.00$^{+1.41}_{-0.00}$ & 0.00$^{+1.13}_{-0.00}$  & 0.00$^{+1.13}_{-0.00}$ & 0.00$^{+1.45}_{-0.00}$\\
0.4  $\leq$  $z$ $<$ 0.5 &  0.00$^{+1.83}_{-0.00}$ & 1.35$^{+2.77}_{-0.41}$ & 0.00$^{+1.14}_{-0.00}$  & 0.00$^{+1.13}_{-0.00}$ & 1.84$^{+3.13}_{-0.62}$\\
0.5  $\leq$  $z$ $<$ 0.6 &  1.74$^{+2.94}_{-0.63}$ & 0.00$^{+1.45}_{-0.00}$ & 0.00$^{+1.16}_{-0.00}$  & 0.00$^{+1.13}_{-0.00}$ & 1.98$^{+3.12}_{-0.72}$\\
0.6  $\leq$  $z$ $<$ 0.7 &  3.31$^{+3.28}_{-1.05}$ & 0.00$^{+1.44}_{-0.00}$ & 0.00$^{+1.19}_{-0.00}$  & 0.00$^{+1.13}_{-0.00}$ & 3.58$^{+3.45}_{-1.13}$\\
0.7  $\leq$  $z$ $<$ 0.8 &  2.17$^{+3.13}_{-0.75}$ & 0.00$^{+2.00}_{-0.00}$ & 0.00$^{+1.29}_{-0.00}$  & 1.00$^{+2.28}_{-0.28}$ & 3.98$^{+3.72}_{-1.29}$\\
0.8  $\leq$  $z$ $<$ 0.9 &  1.26$^{+3.02}_{-0.46}$ & 1.59$^{+2.95}_{-0.56}$ & 0.85$^{+2.42}_{-0.26}$  & 0.00$^{+1.13}_{-0.00}$ & 4.07$^{+3.78}_{-1.39}$\\
0.9  $\leq$  $z$ $<$ 1.0 &  2.94$^{+3.21}_{-0.95}$ & 0.72$^{+3.09}_{-0.14}$ & 0.19$^{+1.85}_{-0.19}$  & 0.00$^{+1.28}_{-0.00}$ & 4.89$^{+4.00}_{-1.56}$\\
1.0  $\leq$  $z$ $<$ 1.1 &  0.00$^{+1.31}_{-0.00}$ & 0.00$^{+2.09}_{-0.00}$ & 0.10$^{+1.85}_{-0.10}$  & 0.00$^{+1.61}_{-0.00}$ & 1.56$^{+3.37}_{-0.59}$\\
1.1  $\leq$  $z$ $<$ 1.2 &  1.05$^{+2.37}_{-0.30}$ & 0.00$^{+1.92}_{-0.00}$ & 0.00$^{+1.15}_{-0.00}$  & 0.00$^{+1.46}_{-0.00}$ & 1.74$^{+3.09}_{-0.57}$\\
1.2  $\leq$  $z$ $<$ 1.3 &  1.03$^{+2.34}_{-0.29}$ & 0.00$^{+1.69}_{-0.00}$ & 0.00$^{+1.13}_{-0.00}$  & 0.00$^{+1.17}_{-0.00}$ & 1.36$^{+2.78}_{-0.41}$\\
1.3  $\leq$  $z$ $<$ 1.4 &  1.00$^{+2.28}_{-0.28}$ & 0.00$^{+1.66}_{-0.00}$ & 1.00$^{+2.28}_{-0.28}$  & 0.91$^{+2.29}_{-0.27}$ & 3.27$^{+3.15}_{-1.00}$\\
1.4  $\leq$  $z$ $<$ 1.5 &  0.00$^{+1.13}_{-0.00}$ & 0.00$^{+1.59}_{-0.00}$ & 0.00$^{+1.13}_{-0.00}$  & 0.00$^{+1.13}_{-0.00}$ & 0.00$^{+1.59}_{-0.00}$\\
1.5  $\leq$  $z$ $<$ 1.6 &  0.00$^{+1.13}_{-0.00}$ & 0.00$^{+1.44}_{-0.00}$ & 0.00$^{+1.13}_{-0.00}$  & 0.00$^{+1.13}_{-0.00}$ & 0.00$^{+1.44}_{-0.00}$\\
1.6  $\leq$  $z$ $<$ 1.7 &  0.00$^{+1.13}_{-0.00}$ & 0.00$^{+1.27}_{-0.00}$ & 0.00$^{+1.13}_{-0.00}$  & 0.00$^{+1.13}_{-0.00}$ & 0.00$^{+1.27}_{-0.00}$\\
     \hline\hline
   \end{tabular}
 \end{center}
\normalsize
\caption[]
{\label{tab:obscount}
The best estimate (\emph{i.e.}, the most probable) number of Ia's, $d_j^m$, in $\Delta z$ = 0.1 redshift bins ($j$ = [1,..,17]), for the 
four samples listed in Table~\ref{tab:samples} ($m$ = [1,..,4]).  
The total numbers are the results of applying the counting 
procedure described in the text to the combined candidates from all four samples
(in other words, the total probability distribution is not a trivial sum of the probability distributions for the four samples).  
All the uncertainties reflect a 68\% confidence region.  
}
\end{table}

\subsubsection{Sensitivity to Varying Priors}
\label{sec:priors}
As usual in Bayesian analysis, the errors on the observed number of supernovae $d_j$
calculated as described in section~\ref{sec:typeia} are a combination of
statistical and systematic uncertainties.  
However, to gain an appreciation for the effect of the prior assumptions
on the final result, we compute the change in $d_j$'s by varying
the calculation of $P({\rm Ia}|\{D_i\},\,z)$ from Eqn.~\ref{eqn:bayes} in three
different ways: 
\begin{itemize}
\item
  {\bf Large extinction}: In section~\ref{sec:typeia}, we considered three discrete cases
for extinction: no extinction, ($A_V$, $R_V$) = (0.4, 2.1), and ($A_V$, $R_V$) = (0.4, 3.1).
We now add the case of ($A_V$, $R_V$) = (1.0, 3.1) to the extinction prior, 
and consider it to  be equally likely as the cases of no extinction and 
moderate extinction.  It is in fact known that a value of $A_V$ = 1.0 is 
much less likely than, say, an $A_V$ = 0; however, it is in cases of strong extinction
that the overlap between the magnitude phase space of Ia's and CC's becomes the largest.
\item
  {\bf Overluminous Ibc's}:  In~\cite{bib:rich2006}, it is pointed out that there may exist
a sub-class of type Ibc supernovae whose mean restframe $B$-band magnitudes are much closer to those
of normal Ia's, with $\bar{M}$ = -20.08, $\delta M$ = 0.46.  We add these supernovae
as one more type to our list of supernova types considered, assuming that $f_{Ibc}$  = 0.18
for normal Ibc's, and 0.09 for the overluminous ones.
\item
  {\bf Flat ratio of the CC to Ia rates, all CC types equally likely}:  
Instead of using the redshift-dependent models for the ratio of the CC to Ia supernova rates, 
we now assume that the ratio is redshift-independent, and taken 
to be 2.15, which is roughly the average of the models shown in Fig.~\ref{fig:models}).
We also assume that the relative fractions of the CC supernovae are all 0.25 ($f_{CC}$ = 0.25);
or, in other words, that all classes of the CC supernovae are equally likely.
\end{itemize}

The considered alternative 
priors are deliberately taken to be such that the effect on $P({\rm Ia}|\{D_i\},\,z)$
should be the most dramatic, without too much regard for whether or not such priors
are realistic.
Table~\ref{tab:systematics} lists the changes in $d_j$ relative to the values
specified in Table~\ref{tab:obscount} as a result of using the alternative priors listed above.
\begin{table}[htbp]
\small
  \begin{center}
    \begin{tabular}{|c|c|c|c|c|c|}
    \hline\hline
    Redshift bin                  & $A_V = 1$ & Overluminous & Flat CC/Ia rates \\
                                  &           & Ibc's        &  $f_{\rm CC}$ = 0.25 \\
\hline
0.0  $\leq$  $z$ $<$ 0.1 &    0.00 &   0.00  &   0.00\\
0.1  $\leq$  $z$ $<$ 0.2 &    0.00 &   0.00  &   0.00\\
0.2  $\leq$  $z$ $<$ 0.3 &    0.00 &   0.00  &   0.00\\
0.3  $\leq$  $z$ $<$ 0.4 &    0.00 &   0.00  &   0.00\\
0.4  $\leq$  $z$ $<$ 0.5 &    0.00 &  -0.14  &  -0.32\\
0.5  $\leq$  $z$ $<$ 0.6 &    0.29 &   0.93  &  -0.24\\
0.6  $\leq$  $z$ $<$ 0.7 &   -0.09 &  -0.86  &  -0.21\\
0.7  $\leq$  $z$ $<$ 0.8 &   -0.16 &  -0.05  &  -0.08\\
0.8  $\leq$  $z$ $<$ 0.9 &   -0.04 &  -0.06  &  -0.61\\
0.9  $\leq$  $z$ $<$ 1.0 &    0.06 &  -0.12  &  -0.56\\
1.0  $\leq$  $z$ $<$ 1.1 &    0.14 &  -0.14  &  -0.03\\
1.1  $\leq$  $z$ $<$ 1.2 &    0.12 &   0.09  &   0.01\\
1.2  $\leq$  $z$ $<$ 1.3 &   -0.01 &  -0.02  &   0.00\\
1.3  $\leq$  $z$ $<$ 1.4 &    0.09 &  -0.08  &   0.04\\
1.4  $\leq$  $z$ $<$ 1.5 &    0.00 &   0.00  &   0.00\\
1.5  $\leq$  $z$ $<$ 1.6 &    0.00 &   0.00  &   0.00\\
1.6  $\leq$  $z$ $<$ 1.7 &    0.00 &   0.00  &   0.00\\
   \hline\hline
   \end{tabular}
 \end{center}
\normalsize
\caption[]
{\label{tab:systematics}
A change in the estimates for the numbers of Ia's, $d_j$, as a result
of using alternative priors for the $\vec{\theta}$ parameters, as described
in the text.  Listed are the differences between the $d_j$ obtained
for the alternative parameters and for the default priors, in $\Delta z$ = 0.1 redshift bins ($j$ = [1,..,17]),
for the combination of the four samples listed in Table~\ref{tab:samples}. 
}
\end{table}

It is clear from Table~\ref{tab:systematics}
that none of the alternative priors considered
leads to a change in the mean that goes beyond the estimated errors  
in Table~\ref{tab:obscount}.

\section{The Rates Calculation}
\label{sec:analysis}
Next, we compute the expected number of candidates
in the $j^{\rm th}$ redshift bin whose center is ${\bar z_j}$,
given a  volumetric  Type Ia supernova rate in the supernova rest frame, 
${\rm r_{V,Ia}({\bar z}) }$,
as a function of redshift $\bar{z}$.  The expected number of candidates
is different from the \emph{measured} $d_j$'s: it is \emph{calculated}
entirely based on Monte Carlo simulations of SNe and a given  rates model.
\begin{equation}
d_j^{\rm exp} = \Delta \bar{z}_j \frac{r_{V,\rm Ia}({\bar z_j})}{1+{\bar z_j}} \, \frac{\Theta}{4\,\pi}\, \frac{dV}{d\bar{z}}({\bar z_j}) \left [T_{\rm Ia} ({\bar z_j}) \epsilon_{\rm Ia}({\bar z_j}) + \frac{r_{V,\rm CC}({\bar z_j})}{r_{V,\rm Ia}({\bar z_j})} T_{\rm CC}({\bar z_j}) \epsilon_{\rm CC}({\bar z_j}) \right],
\label{eqn:rates}
\end{equation}
where $\Delta \bar{z}_j$ is the width of the redshift bin; $\Theta$ is the survey area covered; 
$dV/d\bar{z}$ is the comoving volume
computed assuming a $\Lambda$CDM cosmology with $\Omega_{\Lambda}$ = 0.7,
$\Omega_{M}$ = 0.3, and $H_0$ = 100 $\times$ h$_{70}$ (km/s) Mpc$^{-1}$;
$T_{\rm Ia}(\bar z)$ and $T_{\rm CC}(\bar z)$ are  the control times for 
Ia and non-Ia candidates, respectively; 
$\epsilon_{\rm Ia}$ and $\epsilon_{CC}$ are
the efficiencies of the stage 3 selection for Ia and non-Ia candidates, respectively;
and $r_{V,\rm CC}(\bar z)/{r_{V,\rm Ia}(\bar z)}$ is the ratio of the non-Ia
supernova rate to the Ia supernova rate.  Once again, the appearance of this ratio
does not bias our results, since we do not make any assumptions
about the \emph{absolute} Type Ia rate.
The control time in Eqn.~\ref{eqn:rates} enters with a factor of (1 + $\bar z$).
This is a consequence of the fact that it is calculated in the observer frame,
as will be described later.

The control time $T$ is defined as the time during which a supernova search is
\emph{potentially} capable of finding supernova candidates.
In order to calculate it, we simulate HST observations of 
Type Ia and non-Type Ia supernovae at redshifts up to 1.7,
with the same sampling and exposure times as those of the real data.  By 
shifting the observing grid along the light curves, we calculate the weighted 
sum of the number of days during which a given supernova could be detected.
The weight factors are obtained from the stage 1 efficiency parametrization;
it is also required that the light curves satisfy the stage 2 SNR requirements.
Therefore, stage 1 and 2 supernova selection efficiencies are 
naturally built into the control time calculation.  However, the stage 3 
selection efficiency is  \emph{not} part of the control time calculation, and
must therefore be computed separately.
Calculating the area of the survey is straightforward using a Monte Carlo 
approach.  The calculation of the control time and the survey area
is given in section~\ref{sec:ctandarea}.

The stage 3 selection efficiencies $\epsilon_{\rm Ia}$
and $\epsilon_{\rm CC}$
must be calculated for the candidates that passed the control time requirements,
and thus satisfy both stage 1 and 2 cuts.   We create a Monte Carlo 
sample simulating real supernova candidates of five different types, and apply stage 1 and 2 cuts to them.
We simulate both Ia and non-Ia candidates
and calculate the number of candidates as we would for real data.
This procedure is described in detail in section~\ref{sec:calcnum}.

The errors on the expected $d_j^{\rm exp}$ are a combination of 
statistical and systematic uncertainties.
Apart from the uncertainties inherent in the calculation of
$P({\rm Ia}|\{D_i\},z)$,
the dominant systematic uncertainties come from two sources:
estimating the variation in the control time for Ia's for 
values of the lightcurve timescale stretch, $s$, other than 1,
and estimating the effect of varying the 
ratio of the rates $r_{V,\rm CC}(\bar z)/{r_{V,\rm Ia}(\bar z)}$.
The former is described in more detail in~\ref{sec:ctandarea}, and
for the latter we use two models described in section~\ref{sec:typeia},
for $\tau$ = 1 Gyr and 3 Gyr.


\subsection{The Control Time and Search Area Calculation}
\label{sec:ctandarea}
Let us start with describing the calculation of the control time 
and search area, $T$ and $\Theta$ from Eqn.~\ref{eqn:rates}.
The control time is the time during which a supernova search is
in principle capable of finding supernova candidates on the area
covered.  For the GOODS fields, the orientation of the tiles is such that
a candidate is not necessarily accessible for every search epoch due to edge effects (see Fig.~\ref{fig:goods}).
For example, for sample 1 from  Table~\ref{tab:samples}, 
a given location may only be covered by epochs 1, 3, and 5 (but not by epochs 2 and 4) 
of the North GOODS dataset.  In both our control time calculation and in the search 
area calculation, we thus consider all of the  possible epoch permutations 
at each location: 31 possible permutations for samples 1, 3, and 4 and 15 possible permutations for sample 2.

We perform separate control time and search area calculations for the four samples
listed in Table~\ref{tab:samples}; however, the approach is the same.
For the control time calculation, we make use of 
the simulation described in some detail in Appendix A of~\cite{bib:ourpaper}.
We use it to create simulated HST observations in both F775W and F850LP bands
for Type Ia supernovae of stretch 1, as well as for non-Type Ia 
supernovae, at redshifts up to 1.7 with an increment of 0.1.  
Separate sets of observations are generated for each possible permutation
of the available search epochs, for each of the four samples.  
For example, for a supernova from Sample 1 that happens to be present in 
every one of the North GOODS epochs, there will be five simulated search observations and 
a single reference observation.
We use typical epoch separations and exposure times for a given sample.
The observations are realized using an aperture exposure time calculator with a 0.1'' radius.  
We initially set the explosion date of the supernova
on the last date of the available search epoch observation set (\emph{e.g.},
for the supernova example mentioned above it would be on the date the last of the North
GOODS data were taken).  The observing grid for the search observations is then shifted by one day, and 
the procedure is repeated $N_{shifts}$ = 350 times (that is, spanning approximately
a year, which is the longest separation between the search and reference data 
for our data samples).  
For each such shift, we require that 
the simulated data satisfy both the stage 1 and stage 2 requirements listed in Table~\ref{tab:selection}.  
The resulting control time thus has stage 1 and 2 efficiencies automatically included.   It is given by:
\begin{equation}
T = \sum_{k=1}^{N_{shifts}}  \left[ 1 - \prod_{i = 1}^{N_{ep}}(1-\epsilon_i^k(SNR_i)) \right] e_k,
\end{equation}
where the sum is over all the shifts, $N_{ep}$ is the number of available search epochs (in the
 example considered above, $N_{ep}$ = 5);
$\epsilon_i^k$ is a function of the $i^{\rm th}$  subtraction's SNR, parametrized
as in Eqn.~\ref{eqn:eff}; and $e_k$ is a binary quantity
\begin{eqnarray}
e_k = \left\{
\begin{array}{ll}
 1 & \mbox{, if $k^{\rm th}$ shift configuration satisfies stage 2 requirements}\\
 0 & \mbox{, if $k^{\rm th}$ shift configuration does not satisfy stage 2 requirements}\\
\end{array}
\right.
\end{eqnarray}
that assesses whether a given configuration has enough epochs with sufficient SNR for
the stage 2 selection.

We repeat the control time calculation for Ia's with the lightcurve timescale stretch
values of $s$ = 0.65 and $s$ = 1.30, weight the results by the probability 
of obtaining such stretches taken from Eqn.~\ref{eqn:sprioria}, and take
the larger error between the control time computed for these stretch parameters and 
that computed for a stretch of 1 as a measure of the systematic error on the control time
for Ia's.    
For reference, Table~\ref{tab:ct} lists the control time as a function of redshift for both the nominal
stretch of 1 and for the stretch of 0.65 and 1.30, for the configurations
in which a supernova candidate is assumed present on all of the search epochs.
\begin{table}[htbp]
\small
  \begin{center}
    \begin{tabular}{|c|c|c|c||c|c|c||c|c|c||c|c|c|}
    \hline\hline
                   & \multicolumn{12}{c|}{Control time (yrs)}\\
           \cline{2-13}
  $\bar z$ & \multicolumn{3}{c||}{Sample \#1} & \multicolumn{3}{|c||}{Sample \#2} & \multicolumn{3}{|c||}{Sample \#3} & \multicolumn{3}{|c|}{Sample \#4} \\
                          \cline{2-13}
              & $s$=1 & $s$=0.65 & $s$=1.3 & $s$=1 & $s$=0.65 & $s$=1.3 &$s$=1 & $s$=0.65 & $s$=1.3 &$s$=1 & $s$=0.65 & $s$=1.3 \\ 
     \hline\hline
0.1 &   0.84 &   0.78 &   0.84 &   0.84 &   0.66 &   0.85 &   0.68 &   0.65 &   0.68 &   0.30 &   0.27 &   0.35\\
0.2 &   0.84 &   0.77 &   0.84 &   0.84 &   0.65 &   0.85 &   0.68 &   0.63 &   0.67 &   0.32 &   0.29 &   0.42\\
0.3 &   0.85 &   0.77 &   0.85 &   0.83 &   0.64 &   0.85 &   0.68 &   0.63 &   0.67 &   0.31 &   0.28 &   0.37\\
0.4 &   0.84 &   0.75 &   0.83 &   0.80 &   0.61 &   0.84 &   0.67 &   0.60 &   0.66 &   0.32 &   0.28 &   0.36\\
0.5 &   0.84 &   0.73 &   0.84 &   0.77 &   0.59 &   0.84 &   0.67 &   0.58 &   0.66 &   0.32 &   0.28 &   0.36\\
0.6 &   0.83 &   0.72 &   0.84 &   0.73 &   0.56 &   0.82 &   0.66 &   0.57 &   0.65 &   0.33 &   0.28 &   0.37\\
0.7 &   0.81 &   0.68 &   0.84 &   0.69 &   0.53 &   0.79 &   0.63 &   0.53 &   0.64 &   0.32 &   0.29 &   0.37\\
0.8 &   0.78 &   0.64 &   0.83 &   0.64 &   0.51 &   0.75 &   0.59 &   0.49 &   0.63 &   0.33 &   0.29 &   0.37\\
0.9 &   0.72 &   0.59 &   0.80 &   0.57 &   0.46 &   0.66 &   0.53 &   0.43 &   0.58 &   0.33 &   0.28 &   0.37\\
1.0 &   0.68 &   0.57 &   0.75 &   0.54 &   0.43 &   0.62 &   0.48 &   0.36 &   0.54 &   0.33 &   0.28 &   0.36\\
1.1 &   0.65 &   0.55 &   0.72 &   0.51 &   0.40 &   0.59 &   0.46 &   0.32 &   0.49 &   0.32 &   0.27 &   0.37\\
1.2 &   0.63 &   0.53 &   0.69 &   0.48 &   0.35 &   0.56 &   0.42 &   0.27 &   0.47 &   0.32 &   0.26 &   0.36\\
1.3 &   0.60 &   0.49 &   0.67 &   0.46 &   0.30 &   0.55 &   0.38 &   0.22 &   0.45 &   0.32 &   0.25 &   0.36\\
1.4 &   0.57 &   0.42 &   0.64 &   0.44 &   0.23 &   0.53 &   0.37 &   0.17 &   0.43 &   0.31 &   0.20 &   0.36\\
1.5 &   0.56 &   0.36 &   0.62 &   0.39 &   0.15 &   0.48 &   0.32 &   0.13 &   0.40 &   0.31 &   0.17 &   0.35\\
1.6 &   0.51 &   0.24 &   0.59 &   0.34 &   0.09 &   0.44 &   0.27 &   0.09 &   0.37 &   0.26 &   0.11 &   0.34\\
1.7 &   0.47 &   0.17 &   0.56 &   0.21 &   0.04 &   0.38 &   0.25 &   0.06 &   0.33 &   0.25 &   0.07 &   0.31\\
     \hline\hline
   \end{tabular}
 \end{center}
\normalsize
\caption[]
{\label{tab:ct}
The Type Ia control time in years as a function of redshift, for the configurations on which a
supernova candidate is assumed to be present on all of the search epochs.
The control time is given for three different values of the stretch parameter $s$: 1 (nominal), 0.65, and 1.30.
Note that this control time has the stage 1 and 2 efficiencies built into the calculation.
}
\end{table}

Calculating the search area is non-trivial because of the complicated orientations of the 
GOODS tiles, as well as the overlaps between the tiles (see Fig.~\ref{fig:goods}).  In addition, the search area must be
calculated separately for all of the possible epoch configurations, as described above.
We perform this calculation using a Monte Carlo method.  
First, we create a 300x300 point 
grid between the minimum and maximum right ascensions ($\alpha$) and declinations 
($\delta$) covering the entire North or South GOODS area (\emph{e.g.},
from $\alpha$ = 12:35:34.85 and $\delta$ =  62:4:59.45 
to $\alpha$ = 12:38:14.7 and $\delta$ = 62:23:36.78 for epochs 1, 3, and 5
of sample \# 1).
Then, for a given epoch, and for each
point $i$ on the grid, we check whether this ($\alpha_i$, $\delta_i$) belongs to any of the images
that were used to make subtracted data for this epoch.  In other words, we convert
($\alpha_i$, $\delta_i$) into image coordinates ($x_j$, $y_j$), and check 
that: (a) the point falls within the confines of at least one search/reference image pairs;
(b) it does not fall into the gap between the two ACS chips on the search image; and 
(c) it does not fall on a known bad pixel or a pixel that has been masked off for any other
reason (\emph{e.g.}, due to a residual cosmic ray contamination) on either image,
although because of the drizzling there are very few affected pixels.
If all of these requirements are satisfied, the point is counted toward the 
area calculation.  Once counted, a given point can never again
be counted for this particular epoch.  This avoids double-counting, an issue particularly
important since most GOODS tiles overlap at least somewhat with their immediate neighbors,
and a point with a given ($\alpha_i$, $\delta_i$)  may well be present on several images.
A separate accounting of the number of points is kept for each epoch
permutation.
For example, let us suppose that the number of points that cover all five
of the GOODS North epochs is $a_1$, and that the number of total points tried in the grid
is $A_1$; then the area corresponding to this configuration is $S \, a_1/A_1$, where $S$ is 
the area of the entire North GOODS survey.

\begin{figure}[!htb]
 \begin{center}
    \includegraphics[height=4.in, width=4.8in]{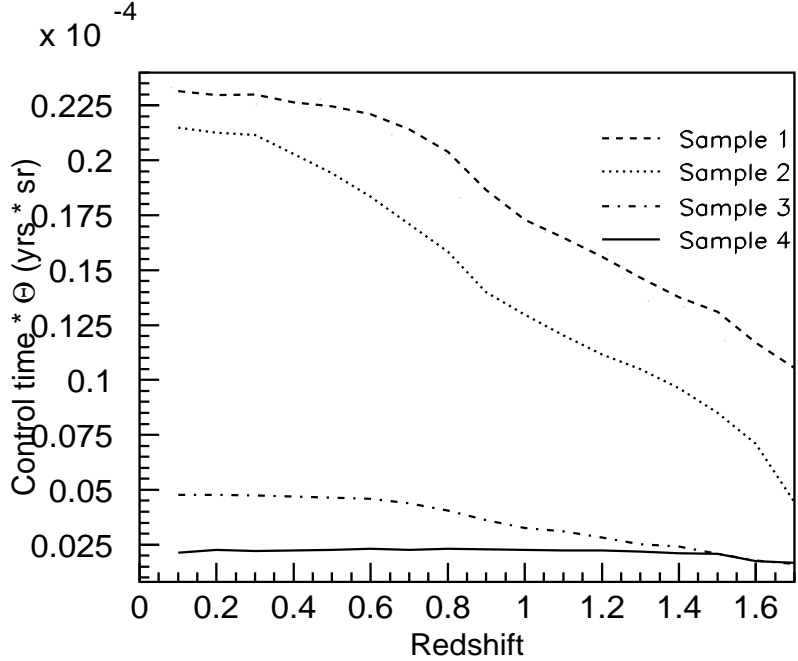}
  \end{center}
\caption[]
{\label{fig:ct}
The product of the control time and surveyed area as a function of redshift for the four
samples listed in Table~\ref{tab:samples}, calculated for a stretch 1  Ia.
The dashed line is for sample \#1,
where the 5 GOODS North epochs were used as the search data, and the combined 2004 ACS sample,
as the reference.  The dotted line is for sample \#2, where the 4 epochs of
the 2004 ACS dataset were used as the search data, and the combined GOODS North
data, as the reference.  The dashed-dotted line is for sample \#3, where
all 5 epochs of the GOODS South sample were used as the search data; and 
the combined epochs 4+5 of the South GOODS dataset, as the reference.
The solid line is for sample \#4, where
all 5 epochs of the GOODS South sample were used as the search data; and 
the combined epochs 1+2 of the South GOODS dataset, as the reference.
}
\end{figure}

Figure~\ref{fig:ct} shows the resulting product of the control time and surveyed area 
($\Theta \, T(\bar z)$ = \\$\sum_{i = 1}^{n} \Theta_i \, T_i(\bar z)$, where $n$ 
is the number of all possible permutations) for stretch 1 Type Ia's,
as a function of redshift for the four different samples in Table~\ref{tab:samples}.
There are several interesting features in Fig.~\ref{fig:ct}.  First, 
the product of the control time and area tends to  decrease with redshift.
This is a consequence of the fact that it becomes more difficult to satisfy
the stage 2 SNR requirements for higher redshift (dimmer) supernovae.
Second, for a given redshift, the product is smaller 
for sample \#2 than for sample \#1, a consequence of the fact that there are only
4 search epochs in sample \#2 vs. 5 search epochs in sample \#1.  
Third, the product is distinctly smaller for the South GOODS
samples (samples \#3 and \#4) than for either of the North samples (\#1 and \#2), a
reflection of the fact that for these samples we are forced to use references made from two of 
the GOODS South dataset's own epochs.  Finally, the product is smaller
for sample \#4, which uses epochs 1+2 of the GOODS South dataset as its reference data,
than for it is for sample \#5, which uses epochs 4+5.  This is simply because
the rise time of a supernova is smaller than its decline time.

\subsection{Calculating $\epsilon_{\rm Ia}$ and $\epsilon_{\rm CC}$}
\label{sec:calcnum}
  In order to determine the efficiency of the stage 3 selection, we generate
four Monte Carlo datasets simulating the data from the four datasets listed in Table~\ref{tab:samples}
(in other words, they have the same sampling, exposure times, \emph{etc.}, as the data).
Each Monte Carlo dataset contains 500 candidates for each of the 5 supernova
types considered (Ia, Ibc, IIL, IIP, and IIn).     
The redshifts of these candidates are drawn from a Gaussian distribution that 
uses the redshifts and redshift errors of the real data events; 
the exposure times and sampling intervals also mimic those 
of the real data.   The candidates' rest-frame 
$B$-band magnitudes, stretch (for Type Ia's), and extinction parameters are drawn from 
the appropriate distributions used in Eqns.~\ref{eqn:final_Ia} and~\ref{eqn:final_nonIa}.
The time period between the date of explosion and 
the first observation is randomly drawn from a flat distribution.
In addition, because we are simulating a dataset as it would appear by the time it is
ready for the stage 3 selection, we impose the same selection requirements from stages 1 and 2
on these Monte Carlo events as we do on the real data.

After these Monte Carlo samples are generated,
we calculate the number of candidates in each redshift bin.
Dividing this number by the total number of the generated Ia's
yields the efficiency $\epsilon_{j\, {\rm Ia}}^m$, for 
redshift bin $j$ for
Monte Carlo dataset $m$.   Similarly, the efficiency for non-Type Ia candidates,
 $\epsilon_{j\, {\rm CC}}^m$, is defined as the sum of the probabilities of the
non-Type Ia candidates divided by the total number of all generated non-Type Ia supernovae.
The values of $\epsilon_{j\, {\rm Ia}}^m$'s range from $\sim$10 to 90\%;
and the values of  $\epsilon_{j\, {\rm CC}}^m$'s, $\sim$3 to 50\%,
depending on the redshift bin.

\subsection{Comparison of Expected and Observed Numbers of Supernovae}
\label{sec:thenumbers}
We can now put everything together and compute the expected numbers of supernovae
for a given model of the Type Ia supernova rates using Eqn.~\ref{eqn:rates}. 
We calculate the observed numbers of supernovae for redshifts 
$\bar{z}$ $\leq$ 1.7, as well as the expected numbers of supernovae for 
the two models considered in~\cite{bib:pain}:
a redshift-independent one and one evolving with redshift as a power law.
We perform a least-squares fit of the observed numbers of supernovae
to the predictions for both models.  
We also perform a  maximum likelihood fit and compare the results.
\begin{itemize}
\item {\bf Redshift-independent rate.}
Assuming the rate is flat as a function of redshift, we obtain the
best-fit value of 
${\rm r_{V,Ia}}$  = 
(1.1 $^{+0.2}_{-0.2}$) $\times$ 10$^{-4}$ $N_{\rm Ia}$/(year Mpc$^3$ $h_{70}^{-3}$)
, with a $\chi^2$ = 11.5 for 16 degrees of freedom.
Figure~\ref{fig:numsn} (left) shows the resulting distribution of the predicted
and observed numbers of supernovae.   
The errors on the predicted numbers of supernovae are
a quadratic combination of the statistical and systematic errors (the statistical and 
systematic errors are comparable).
The maximum likelihood method yields  
${\rm r_{V,Ia}}$ = (0.7$^{+0.2}_{-0.2}$) 
$\times$ 10$^{-4}$ $N_{\rm Ia}$/(year Mpc$^3$ $h_{70}^{-3}$),
consistent with the $\chi^2$ method.

\item {\bf Rate evolving as a power law with redshift.}
Assuming the rate is varying as a function of redshift as (1 + $\bar z$)$^\alpha$,
using the $\chi^2$ fitter we obtain the best-fitting value for 
$\alpha$ = 0.2$^{+0.7}_{-0.7}$ with a $\chi^2$ = 11.4 for 15 degrees of freedom.  
This is consistent with~\cite{bib:pain}, who found $\alpha$ = 0.8 $\pm$ 1.6.
Note that the fit results are also consistent with
 the $\alpha$ = 0 case that was considered above.
The maximum likelihood method yields $\alpha$ = -0.4$^{1.0}_{1.1}$.
\begin{figure}[!htb]
 \begin{center}
$\begin{array}{c@{\hspace{0.0in}}c}
\epsfxsize=3.3in\epsfysize=2.8in\epsffile{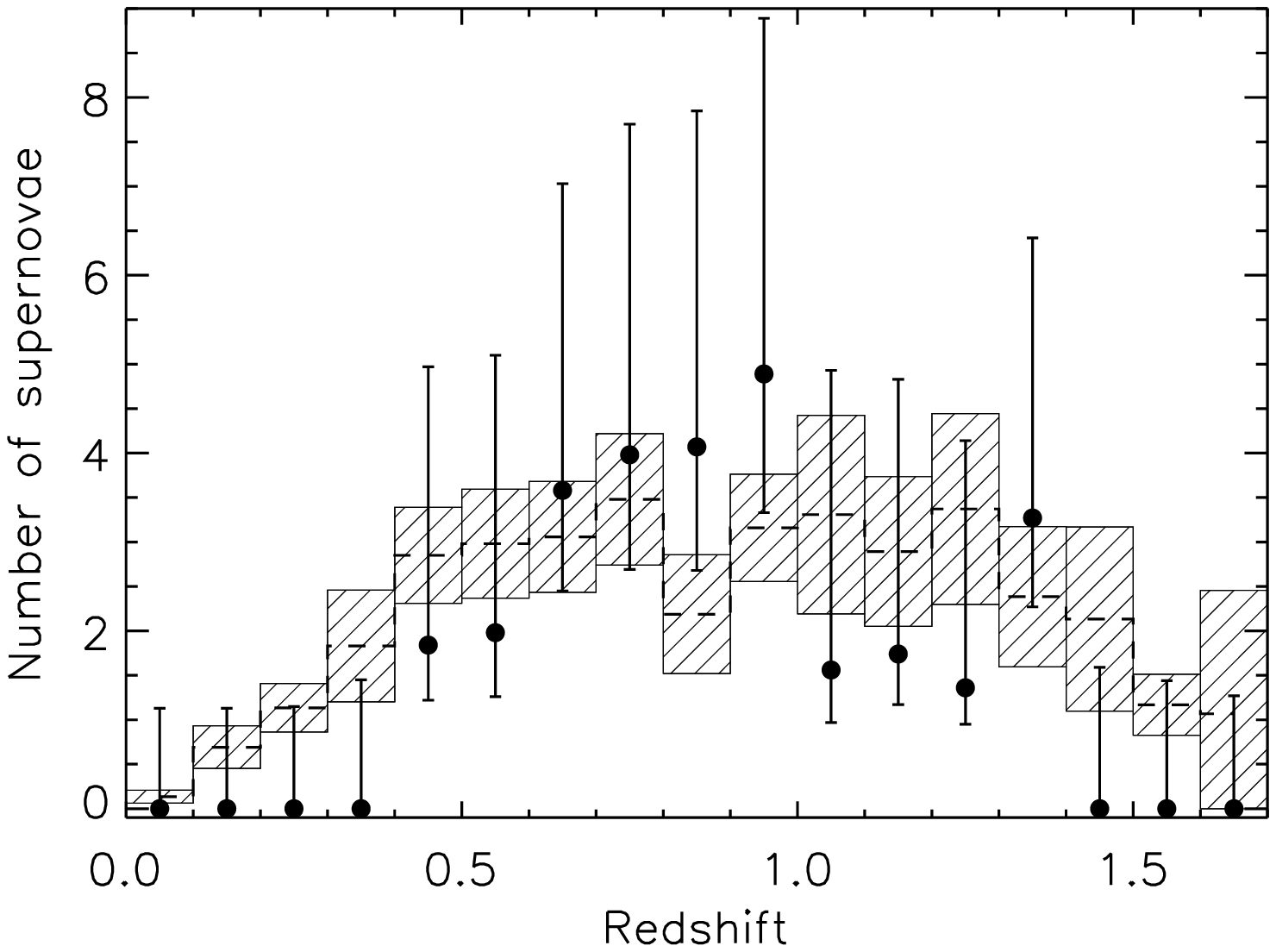} & 
\epsfxsize=3.3in\epsfysize=2.8in\epsffile{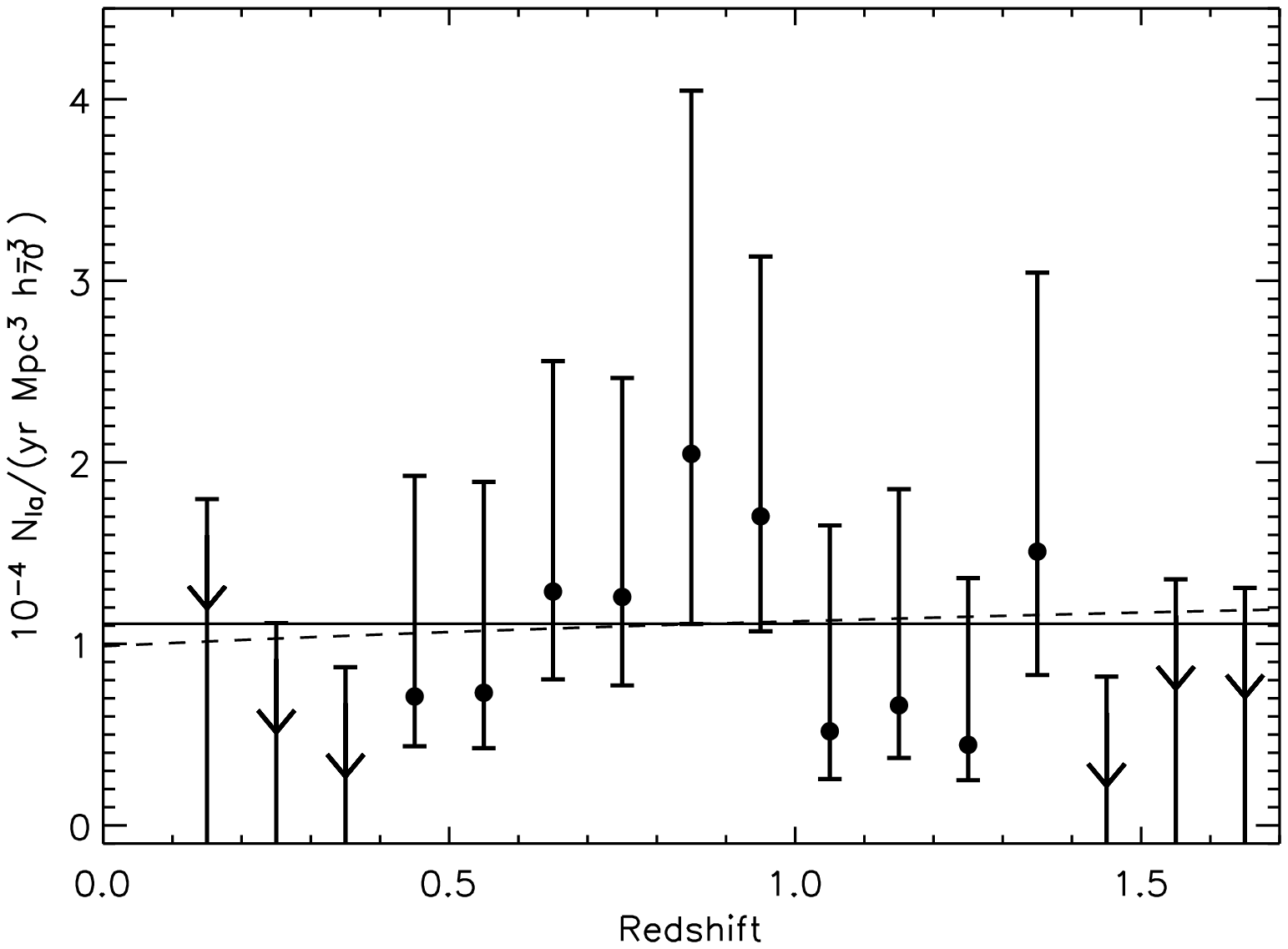} 
\end{array}$
  \end{center}
\caption[]
{\label{fig:numsn}
Left: The total observed candidates for the 4 samples are plotted
as a function of redshift (filled circles).
The errors on the observed candidates are given in Table~\ref{tab:obscount}.
The predicted number of candidates  has been computed assuming
a redshift-independent volumetric Type Ia rate of 
${\rm r_{V,Ia}}(\bar z)$ = 1.1 $\times$ 10$^{-4}$  $N_{\rm Ia}$/(year Mpc$^3$ $h_{70}^{-3}$)]),
and is plotted as a dashed histogram.
The shaded region around the predicted numbers indicates the range of combined statistical
and systematic errors.
The contributions from the statistical and systematic errors are comparable.
Right: The calculated rates as a function of redshift (filled circles),
with overplotted fit results to the fits described in the text:
redshift-independent rate (solid line) and power-law redshift dependent rate (dashed line).
Note that the plot does not show the rates in the first redshift bin;
this is because in this bin the rates are effectively unconstrained on the scale shown.  
}
\end{figure}
\end{itemize}
Both the redshift-independent model and the power-law model yield acceptable
fit results, judging by the obtained  
$\chi^2$'s (note, however, 
that the data points in neighboring bins are correlated, leading to lower $\chi^2$ per DOF).  
The probability $p(\Delta \chi^2|\Delta DOF)$ = 0.1,
where $\Delta \chi^2$ and $\Delta DOF$ are the difference in the $\chi^2$ 
and the numbers of degrees of freedom, respectively, for the redshift-independent model
and the power-law model.
In other words, the approximate probability that data would 
fluctuate from the redshift-independent  model to the power-law model
is 0.1.
This fact indicates that our description of the experiment is
good at both low and high redshifts.  One must note, however, that at 
redshifts $>$ 1 the samples start becoming sparser, and at redshifts 
$>$ ~1.4 the measurement becomes particularly difficult with this dataset.

\section{Comparison to Rates in the Literature}
\label{sec:comp}
To compare our results with those of~\cite{bib:dahlen},
we now compute the Type Ia supernova rates in four large redshift bins,
0.2 $\leq$ $\bar z$ $<$ 0.6, 0.6 $\leq$ $\bar z$ $<$ 1.0, 1.0 $\leq$ $\bar z$ $<$ 1.4, 
and 1.4 $\leq$ $\bar z$ $<$ 1.7.  
Table~\ref{tab:candscount} enumerates the estimates for the number of candidates in 
these redshift bins for  the four samples listed in Table~\ref{tab:samples}.  
\begin{table}[htbp]
  \begin{center}
  \small
    \begin{tabular}{|c|c|c|c|c|c|}
    \hline\hline
       Redshift bin             & $d_j^1$ & $d_j^2$ & $d_j^3$ & $d_j^4$ & Total\\
     \hline\hline
0.2  $\leq$  $z$ $<$ 0.6 &  2.40$^{+3.07}_{-0.86}$  (0) & 1.81$^{+3.05}_{-0.60}$ & 0.00$^{+1.17}_{-0.00}$ (0) & 1.00$^{+2.28}_{-0.28}$ (2) & 5.44$^{+3.90}_{-1.63}$\\
0.6  $\leq$  $z$ $<$ 1.0 &  13.40$^{+8.28}_{-5.22}$ (6) & 3.85$^{+3.47}_{-1.25}$ & 1.71$^{+2.90}_{-0.63}$ (2) & 1.07$^{+2.42}_{-0.30}$ (2) & 18.33$^{+4.62}_{-4.62}$\\
1.0  $\leq$  $z$ $<$ 1.4 &  3.23$^{+3.07}_{-0.97}$  (3) & 2.01$^{+3.25}_{-0.75}$ & 1.50$^{+2.74}_{-0.48}$ (1) & 1.70$^{+2.62}_{-0.58}$ (1) & 8.87$^{+3.13}_{-2.36}$\\
1.4  $\leq$  $z$ $<$ 1.7 &  0.00$^{+1.13}_{-0.00}$  (0) & 0.35$^{+1.72}_{-0.35}$ & 0.00$^{+1.13}_{-0.00}$ (1) & 0.00$^{+1.13}_{-0.00}$ (1) & 0.35$^{+1.72}_{-0.35}$\\
     \hline\hline
   \end{tabular}
 \end{center}
\normalsize
\caption[]
{\label{tab:candscount}
The best estimate (\emph{i.e.}, the most probable) number of Ia's, $d_j^m$, 
in $\Delta \bar{z}$ = 0.4 (0.3 for the last bin) redshift bins ($j$ = [1,..,4]), for the 
four samples listed in Table~\ref{tab:samples} ($m$ = [1,..,4]). 
 The numbers in parenthesis are the number of gold and silver Ia's in the sample from~\cite{bib:riess2},
that were used in the rates analysis of~\cite{bib:dahlen}.
The total numbers are the results of applying the counting 
procedure described in the text to the combined candidates from all four samples
(in other words, the total probability distribution is not a trivial sum of the probability distributions for the four samples).  
All the uncertainties reflect a 68\% confidence region.
}
\end{table}

\begin{figure}[!htb]
 \begin{center}
    \includegraphics[height=4.0in, width=5.0in]{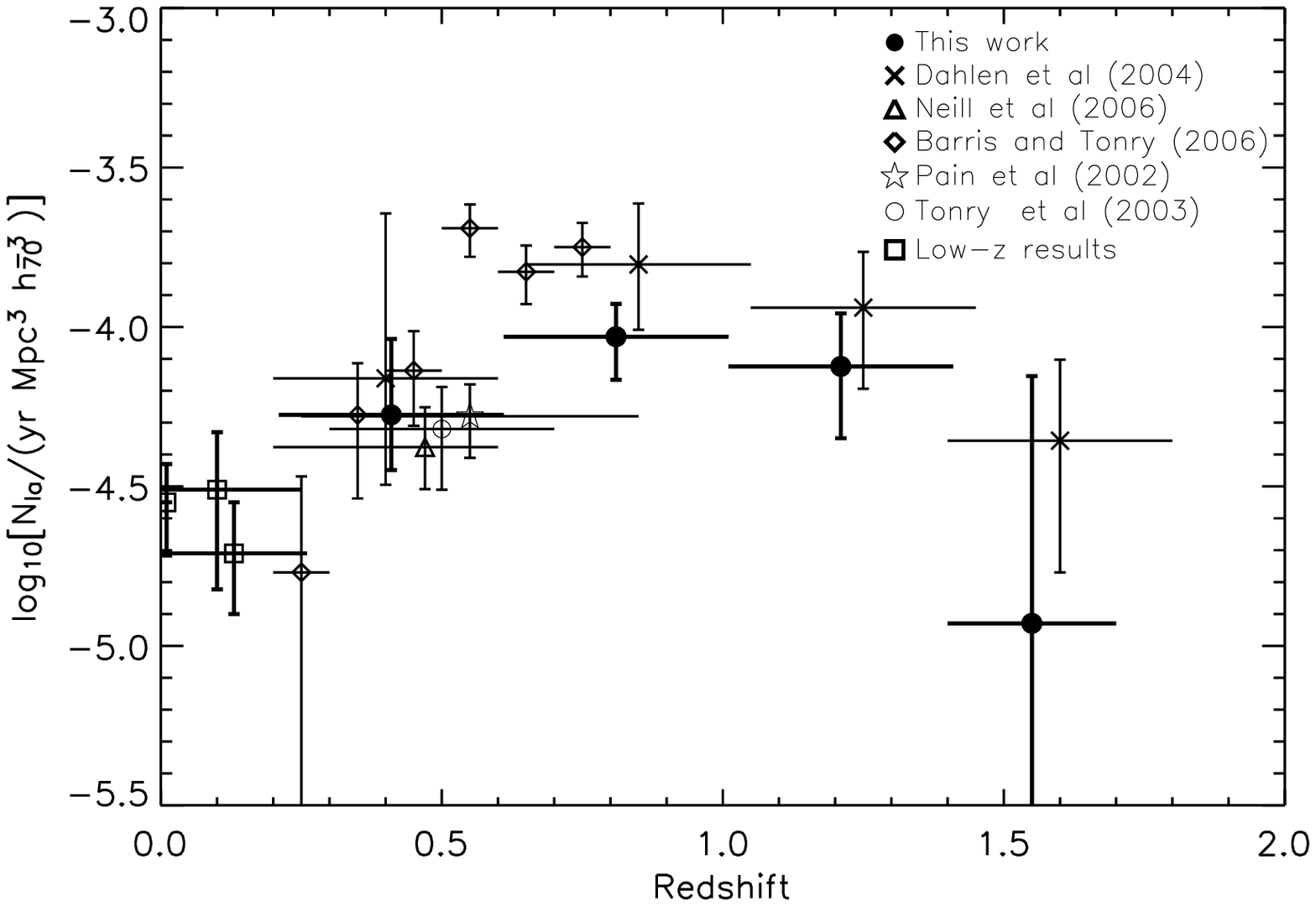}
  \end{center}
\caption[]
{\label{fig:allrates}
The filled circles are the results of this work.
The first three empty squares at low redshifts (the ``low-$z$ results'') are, from lower to higher redshifts,
the results of~\cite{bib:cappellaro},~\cite{bib:madgwick}, and~\cite{bib:blanc}, respectively.  The open upturned triangle at $z$ = 0.47 is from~\cite{bib:neill}.
The open circle at $z$ = 0.50 is from~\cite{bib:tonry}.
The open star at $z$ = 0.55 is from~\cite{bib:pain}.
The open diamonds are the results of~\cite{bib:barris}. 
The crosses are from~\cite{bib:dahlen} (including systematic errors).
The horizontal bars are estimated redshift bin sizes.
}
\end{figure}

Using all four samples, we can now compute the rates for each bin
using Eqn.~\ref{eqn:rates}.
The values for  $\Theta T (\bar z)$, $dV/d\bar{z}$, $r_{V,\rm CC}/r_{V,\rm Ia}$,  and $\bar z$ are taken in the middle of the bin.
The errors on the rates are a quadratic combination of the errors on the number of observed
Ia's listed in Table~\ref{tab:candscount}, as well as statistical and 
systematic errors on the right-hand-side part of Eqn.~\ref{eqn:rates}.
The resulting rates are summarized in Table~\ref{tab:results}
and plotted in Fig.~\ref{fig:allrates} together with
the rates from~\cite{bib:dahlen} and 
results from the literature at lower redshifts.
It is apparent that our results are consistent with those from the literature:
in particular, at higher redshifts our rates are not
inconsistent with those
of~\cite{bib:dahlen}, although obtaining a precise measure of the consistency
would require a careful evaluation of the correlations between the 
samples used in both analyses.
\begin{table}[htbp]
  \begin{center}
    \begin{tabular}{|c|c|}
    \hline\hline
       Redshift  bin        & ${\rm r_{V,Ia}(\bar{z}) }$\\
                            & ([10$^{-4}$ $N_{\rm Ia}$/(year Mpc$^3$ $h_{70}^{-3}$)]) \\
     \hline\hline
       0.2 $\leq$ $\bar z$ $<$ 0.6   &  $0.53^{+0.39}_{-0.17}$ \\
       0.6 $\leq$ $\bar z$ $<$ 1.0   &  $0.93^{+0.25}_{-0.25}$\\
       1.0 $\leq$ $\bar z$ $<$ 1.4   &  $0.75^{+0.35}_{-0.30}$\\
       1.4 $\leq$ $\bar z$ $<$ 1.7   &  $0.12^{+0.58}_{-0.12}$\\
     \hline\hline
   \end{tabular}
 \end{center}
\caption[]
{\label{tab:results}
The Type Ia supernova rates in the four redshift bins considered.  
The errors are a quadratic combination of the errors on 
$d_j^m$'s listed in Table~\ref{tab:candscount}, as well
as statistical and systematic errors on the right-hand-side of 
Eqn.~\ref{eqn:rates}. 
} 
\end{table}

Note that Type Ia supernovae that we have considered encompass a wide range of magnitudes,
stretch parameters, extinction possibilities, etc..   Therefore, 
the procedure described in section~\ref{sec:typeia} accounts for not
only the more standard Type Ia's (such as those described in~\cite{bib:branch}), but also
non-standard Type Ia's, such as type 1991bg and 1991T~\cite{bib:fil}.
1991bg-like supernovae have low values of the stretch parameter 
($s$ = 0.71 $\pm$ 0.05), and are typically $\sim$1.7 magnitudes
fainter in the $V$-band and $\sim$2.6 magnitudes fainter in the $B$-band.
Stretch values of 0.71 are certainly within the range of the stretch parameters
we considered;  as for the magnitudes, it is reassuring to note that the 
case of strong extinction ($A_V$ = 1) did not significantly alter our results 
(see Table~\ref{tab:systematics}).
1991T-like supernovae are about 0.5-0.9 magnitudes brighter
than normal Ia's, with stretch $s$ = 1.07 $\pm$ 0.06.  Both 
the stretch and the magnitude values are well within the considered ranges 
of these parameters.
Note also that the fact that the Bayesian classification method was able to accurately 
type the vast majority of the 73 Type Ia candidates from the SNLS dataset,
as was demonstrated in~\cite{bib:ourpaper}, shows that the method is capable
of identifying Type Ia's in large populations that presumably include 
non-standard Ia's.

It is particularly interesting to compare our rate results with that of~\cite{bib:dahlen}. 
That study also analyzed the GOODS sample, but there are important differences
in our methods, as pointed out above (Sec. 1). While our results are in statistical agreement, 
 our measured rate in a given bin can differ from theirs through either the candidate counting or 
the calculation of the control time/efficiency.
\begin{itemize}
\item
{\bf  Candidate counting: }
In some bins, the  final count of the candidates ends up being about the same for
both analyses, but the actual candidates are not the same. This is not unexpected
because the techniques used for the supernova identification in the two analyses are quite
different. Our method provides a probabilistic rather than an absolute identification of each
individual supernova based on its photometric measurements alone; the same probabilistic
approach is used for calculating the efficiency and mis-identification. 

For example, in the highest redshift bin, 
we have one candidate but this is from Sample 2, which was taken after the 
work of~\cite{bib:dahlen} was published.
However, the two high-redshift Ia candidates from Samples 3 and 4, SN-2002fx and SN-2003ak, 
which were used in~\cite{bib:dahlen}, did not pass our stage 2 cuts.

\item
{\bf  Control time/efficiency:} 
A rigorous comparison of the control times is difficult due to the
lack of tabulated control time data in~\cite{bib:dahlen}. However, a rough estimation of
the control time times efficiency factor from the data given in~\cite{bib:dahlen} shows that
this factor is approximately half our values for all but the highest redshift bin.
\end{itemize}

\section{The Star Formation History Connection}
\label{sec:sf}
A particularly interesting aspect of a Type Ia supernova rates analysis is 
the possibility of constraining the delay time between the formation of 
a progenitor star and a supernova explosion, which in turn helps constrain
possible models for the Type Ia supernova formation.  
There are two leading models that have been considered 
in the recent literature: the so-called two-component model and 
a Gaussian delay model.  We will now consider both of these models.
Unlike section~\ref{sec:thenumbers}, now that we are considering the rates we can add the low-redshift 
measurements of~\cite{bib:cappellaro},~\cite{bib:madgwick}, and~\cite{bib:blanc} to 
our results and use the combined data in the fits.

The two-component model~\citep{bib:scan,bib:mannucci} suggests that that the delay function may be bimodal, 
with one component responsible for the ``prompt'' Type Ia supernovae that
explode soon after the formation of their progenitors; and the other,
for the ``tardy'' supernovae that have a much longer delay time.
Following this model, the Type Ia supernova rate can be represented as:
\begin{equation}
{\rm r_{V,Ia}(\bar{z}) } = A\rho_{\ast} (\bar{z}) + B \dot{\rho}_{\ast}(\bar{z}),
\label{eqn:sfh}
\end{equation}
where $\rho_{\ast} (\bar{z})$ is the integrated SFH and  
$\dot{\rho}_{\ast}(\bar{z})$ is the instantaneous SFH.
The first term of the equation
accounts for the ``tardy'' population, while the second, for the ``prompt'' one.
We use the parametric form of the SFH as given in~\cite{bib:handb}:
\begin{equation}
\dot{\rho}_{\ast}(\bar{z}) = \frac{ (a + b\bar{z}) \, h_{70} }{1 + (\bar{z}/c)^d},
\end{equation}
where $h_{70} = 0.7$, $a$ = 0.017, $b$ = 0.13, $c$ = 3.3, $d$ = 5.3.

\begin{figure}[!htb]
 \begin{center}
  $\begin{array}{c@{\hspace{-0.1in}}c}
  \epsfxsize=3.2in\epsfysize=2.8in\epsffile{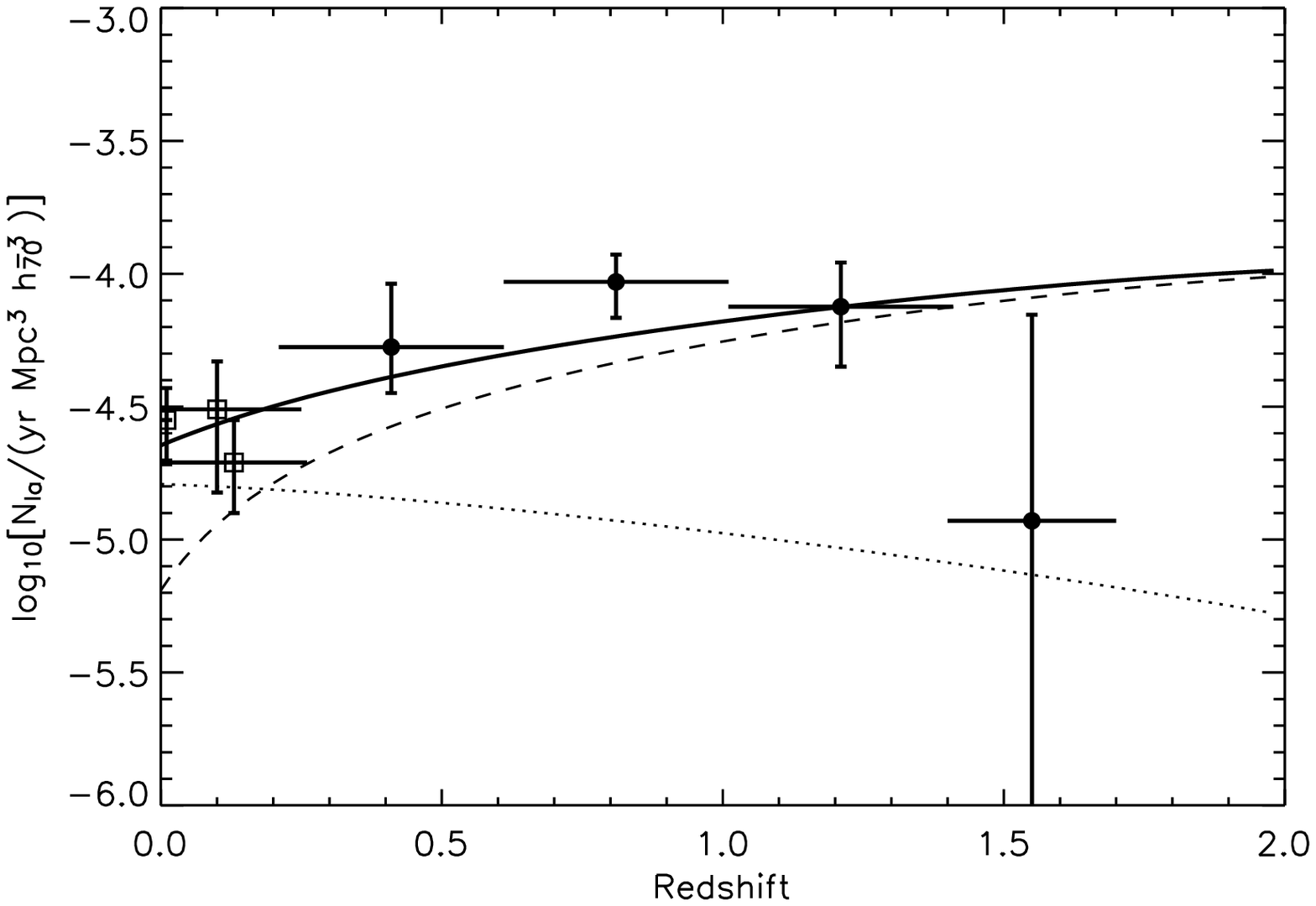} & 
  \epsfxsize=3.2in\epsfysize=2.8in\epsffile{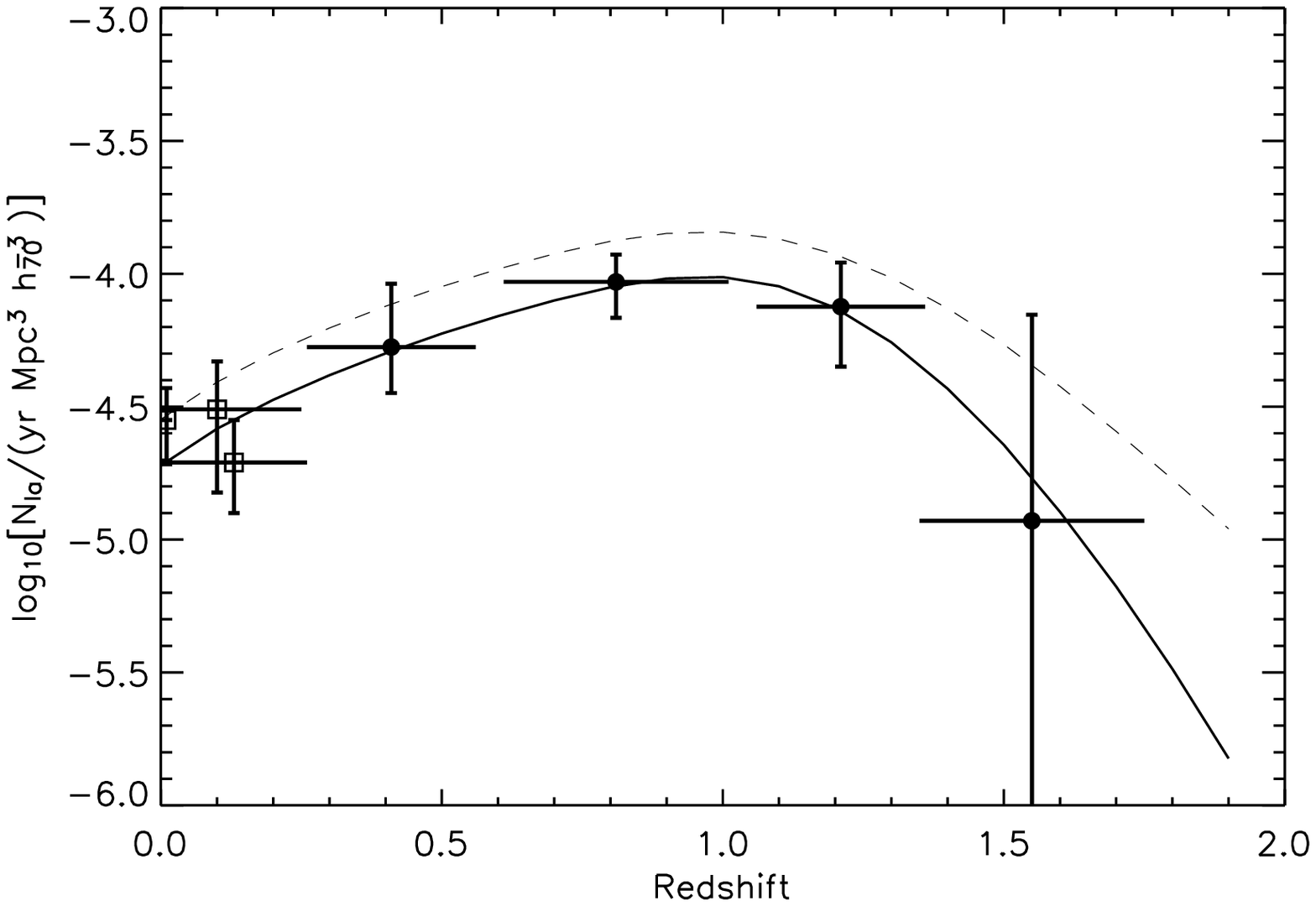}\\
  \end{array}$
  \end{center}
\caption[]
{\label{fig:models}
Left: The least-square fit of the two-component model to the data.
The dashed line represents the prompt component that is proportional to the
instantaneous SFH.  The dotted line represents the tardy component that is
proportional to the integrated SFH.  The thick solid line is the sum of the two.
Right: The Gaussian time delay model with our best-fitted parameters
(solid line), as well as with the parameters of~\cite{bib:strolger} (dashed line).
In both plots, the first three empty squares at low redshifts are, from lower to higher redshifts,
the results of~\cite{bib:cappellaro},~\cite{bib:madgwick}, and~\cite{bib:blanc}, respectively.
The filled circles are the results of this work.
The horizontal bars are estimated redshift bin sizes.
}
\end{figure}

The Levenberg-Marquardt least-squares fit of the 
combined data to the two-component model is shown in Fig.~\ref{fig:models} (left).  We obtain 
$A$ = (1.5 $\pm$ 0.7) 
$\times 10^{-14}$ yr$^{-1}$M$_\odot^{-1}$ and 
$B$ = (5.4 $\pm$ 2.0) 
$\times 10^{-4}$ yr$^{-1}$/(M$_{\odot}$ yr$^{-1}$).
These results are entirely consistent with those
obtained by~\cite{bib:neill}: 
$A$ = (1.4 $\pm$ 1.0) $\times 10^{-14}$ yr$^{-1}$M$_\odot^{-1}$ and 
$B$ = (8.0 $\pm$ 2.6) $\times 10^{-4}$ yr$^{-1}$/(M$_{\odot}$ yr$^{-1}$).
The $\chi^2$ of the fit is 5.4 for 5 degrees of freedom.

Note that the results of~\cite{bib:barris} at  $z$ = 0.55, 0.65, and 0.75
are somewhat inconsistent with our best-fitting two-component model,
with the discrepancy at the level of 4.1, 3.2, and 5.2 $\sigma$, respectively.
This can be seen from Fig.~\ref{fig:allrates}.
It has been argued in~\cite{bib:neill} (who also noted that the results of~\cite{bib:barris}
beyond the redshift of 0.5 appear to be rather high) that contamination by non-Type Ia's
is the most likely source of the problem.

It was suggested in~\cite{bib:dahlen} and~\cite{bib:strolger} that the Ia rate 
is a convolution of the SFH and a Gaussian time delay distribution function with
a characteristic time delay $\tau$ $\sim$3 Gyr and a $\sigma$ = 0.2 $\tau$.
Using the Hopkins-Beacom SFH, we find that the best-fitting
parameters for such a model are 
$\tau$ = (3.2 $\pm$ 0.6) Gyr and $\sigma$ = (0.12 $\pm$ 0.54) $\tau$,
with a fit $\chi^2$ of 2.1 for 4 degrees of freedom.
The fit is shown in Fig.~\ref{fig:models} (right).
For comparison, we also show the rate model obtained using the parameters 
from~\cite{bib:strolger} ($\tau$ = 3 Gyr and $\sigma$ = 0.2 $\tau$).

One of the main differences between the two-component model and the time delay model
is the predicted behavior at high redshifts: the former predicts an increase in the rates,
while the latter, a decrease.  From Fig.~\ref{fig:models} and the results
of the fits of our data to both models, we find that neither scenario
can be ruled out.

\section{Summary}
\label{sec:concl}
We have analyzed the rates of Type Ia supernovae up to a
redshift of 1.7 using two samples
collected with the HST: the GOODS data, and 
the 2004 ACS sample 
collected in the Spring-Summer 2004 covering the GOODS North field.  Using only
the data from two broadband filters, F775W and F850LP, we applied
a novel technique for identifying Type Ia supernovae based on a Bayesian probability
approach.  
This method allows us to automatically type supernova candidates
in large samples, properly taking into account all known sources of systematic
error.  
We also make use of the best currently available full spectral templates for 
five different supernova types for the candidate typing, as well as 
for calculating the efficiency of our supernova search, 
and the control time.  
These templates will undoubtedly be improved over the next several years 
as more supernova data becomes available.  Current and upcoming supernova
surveys will not only provide a better understanding of individual supernova
types, but may also uncover new types of supernovae,
which can then be added to the Bayesian classification framework.
Likewise, a better understanding of the many parameters that affect supernova
observations will improve  the classification scheme, which will
result in better constraints on the measured rates.
The calculations of the supernova finding efficiency, the control time,
and the survey area are all done taking into account the specific observing configurations
pertinent for the surveys, such as exposure times, cadences, and the orientations of the 
GOODS tiles.

We carried out a comparison of the predicted and observed numbers of 
supernovae 
in redshift bins of $\Delta \bar{z}$ = 0.1, for two different models 
of the Type Ia supernova rates:  a redshift-independent rate
and a power-law redshift-dependent rate.
We find that the available data fit both models equally well.

For comparison with previous work, particularly that of~\cite{bib:dahlen}, who 
also analyzed a large subset of the data used here, 
we calculated the volumetric Type Ia 
supernova rates in four redshift bins, 
0.2 $\leq$ $\bar z$ $<$ 0.6, 0.6 $\leq$ $\bar z$ $<$ 1.0, 1.0 $\leq$ $\bar z$ $<$ 1.4, 
and 1.4 $\leq$ $\bar z$ $<$ 1.7.
We find that our results are generally consistent with 
those of~\cite{bib:dahlen}.
Due to the larger of number supernova candidates which this Bayesian  
classification technique makes available, we obtain
smaller or equal uncertainties in all the bins up to $z$ = 1.7. In the highest  
redshift bin we obtain a larger uncertainty because the signal to  
noise ratio is generally too low to apply this technique.

We fitted the resulting rates to two leading models used
in recent literature: the two-component model and a Gaussian time 
delay model.  The former model implies an increase in the Type Ia supernova 
rates at highest redshifts; while the latter, a decrease.  
We find that the statistics of the present sample does not 
definitively discriminate between the two scenarios -- 
only one supernova in this work and two supernovae in~\cite{bib:dahlen}
contribute to the important highest-redshift bin.
Significantly larger surveillance time would be required to 
arrive at a conclusive statement on the trends for the Type Ia rates
at high redshifts.

In the future, several ambitious new surveys are planned that will collect
photometric data for thousands of supernovae in order to improve the
constraints on dark energy.  Individual spectroscopic follow up for every
supernova candidate is likely to be impractical in these surveys.
The Bayesian classification method described here has the ability to 
classify supernovae using 
photometric measurements alone,
and is a promising technique for these future surveys.

\section{Acknowledgments}
We would like to thank Tomas Dahlen and Bahram Mobasher for providing us with photometric
redshifts for a number of supernova candidates.  
We would also like to thank the anonymous referee for the many 
insightful comments and suggestions.
Financial support for this work was provided by NASA through program GO-9727
from the Space Telescope Science Institute, which is operated by 
AURA, Inc., under NASA contract NAS 5-26555.
This work  was also partially supported by the 
Director, Office of Science, Department of Energy, under 
grant DE-AC02-05CH11231.


\end{document}